\documentclass[preprint2]{aastex}
\usepackage{graphicx}
\usepackage{txfonts}
\usepackage{natbib}
\usepackage{longtable} 
\newcommand\gcn{GCN Circ.}

\slugcomment{To appear in ApJ}

\shorttitle{Flares in Swift-detected GRBs}
\shortauthors{Chincarini et al.}

\begin{document}
\title{The First Survey of X-ray Flares from Gamma Ray Bursts Observed by Swift: Temporal Properties and Morphology}
\author{G.\ Chincarini\altaffilmark{1,2}, A.\ Moretti\altaffilmark{1}, 	P.\ Romano\altaffilmark{1,2}, 
	A.D.\ Falcone\altaffilmark{3}, D.\ Morris\altaffilmark{3}, J.\ Racusin\altaffilmark{3}, 
	S.\ Campana\altaffilmark{1}, C.\ Guidorzi\altaffilmark{1}, G.\ Tagliaferri\altaffilmark{1}, 
	D.N.\ Burrows\altaffilmark{3}, C.\ Pagani\altaffilmark{3}, M.\ Stroh\altaffilmark{3}, D.\ Grupe\altaffilmark{3},  
	M.\ Capalbi\altaffilmark{4},	G.\ Cusumano\altaffilmark{5},	N.\ Gehrels\altaffilmark{6}, 
	P.\ Giommi\altaffilmark{4},	V.\ La Parola\altaffilmark{5},	V.\ Mangano\altaffilmark{5},
	T.\ Mineo\altaffilmark{5},	J.A.\ Nousek\altaffilmark{3},	P.T.\ O'Brien\altaffilmark{7},	
	K.L.\ Page\altaffilmark{7},	M.\ Perri\altaffilmark{4},	E.\ Troja\altaffilmark{5}, 
	R.\ Willingale\altaffilmark{7},
	B.~Zhang\altaffilmark{8}
} 
\altaffiltext{1}{INAF--Osservatorio Astronomico di Brera, Via E.\ Bianchi 46, I-23807 Merate (LC), Italy}
\altaffiltext{2}{Universit\`a{} degli Studi di Milano, Bicocca, Piazza delle Scienze 3, I-20126, Milano, Italy}
\altaffiltext{3}{Department of Astronomy \& Astrophysics, Pennsylvania State University, University Park, 
    PA 16802, USA}
\altaffiltext{4}{ASI Science Data Center, Frascati, Italy}
\altaffiltext{5}{INAF- Istituto di Fisica Spaziale e Fisica Cosmica sezione di Palermo, Palermo, Italy}
\altaffiltext{6}{NASA/Goddard Space Flight Center, Greenbelt, MD}
\altaffiltext{7}{Department of Physics and Astronomy, University of Leicester, Leicester, UK}
\altaffiltext{8}{Department of Physics, University of Nevada, Las Vegas, NV 89154-4002, USA}

\begin{abstract}
We present the first systematic investigation of the morphological and
timing properties of flares in GRBs observed by Swift/XRT.
We consider a large sample drawn from all GRBs detected by Swift,
INTEGRAL and HETE-2 prior to 2006 Jan 31, which had an XRT follow-up and which
showed significant flaring. 
Our sample of 33 GRBs includes long and short, at low and
high redshift, and a total of 69 flares. The strongest flares occur in the
early phases,  with a clear anti-correlation between the flare peak intensity
and the flare time of occurrence. Fitting each X-ray flare with
a Gaussian model, we find that the mean ratio of the width and peak
time is $\langle \Delta t / t \rangle = 0.13\pm0.10$, albeit with a large scatter. 
Late flares at times $> 2000$ seconds have long durations,  $\Delta t>300$ s, 
and can be very energetic compared to the underlying continuum.
We further investigated if there is a clear link between the number of pulses
detected in the prompt phase by BAT and the number of X-ray flares 
detected by XRT, finding no correlation.
However, we find that the distribution of intensity ratios
between successive BAT prompt pulses and that between successive XRT flares
is the same, an indication of a common origin for gamma-ray pulses and X-ray flares.
All evidence indicates that flares are indeed related to the
workings of the central engine and, within the standard fireball
scenario, originate from internal shocks rather than external shocks.
While all flares can be explained by long-lasting engine activity, 29/69 flares
may also be explained by refreshed shocks. However, 10 can {\it only}
be explained by prolonged activity of the central engine.

\end{abstract}

	\section{Introduction\label{flares:introduction}}

The advent of Swift \citep{Gehrels04:SWIFT} has brought substantial advances in 
our knowledge of GRBs including the discovery of the first afterglow (with a 
position known to several arcsec precision) of a short burst.  
Swift also brought on the definition of a possible third class of GRBs 
\citep[][]{Gehrels2006:060614Nat}, 
the discovery of a smooth transition between prompt and afterglow emission 
\citep[][]{Tagliaferri2005:nature,Vaughan2006:050315,Obrien2006:xrtbat},  
and the definition of a canonical X-ray light curve 
\citep[][]{Nousek2006:lcvs,Obrien2006:xrtbat,Zhang2006:theory_from_xrt}. 
The latter includes a steep early part ($\propto t^{-\alpha_1}$ with $3\la\alpha_1\la5$,
typically interpreted as GRB high-latitude emission),
a flat phase ($0.5\la\alpha_2\la1$, generally interpreted as due to energy injection into the 
external shock), and a last, steeper part ($1\la\alpha_3\la1.5$, the only one observed by 
pre-Swift X-ray instruments), 
with the predicted $t^{-1}$ decay (see, also \citealt{Wu2006:xflares}). 
Sometimes, a further steepening is detected after the normal decay phase, which
is consistent with a jet break \citep{Zhang2006:theory_from_xrt}. 

What may be the most surprising discovery 
is the presence of flares in a large percentage of X-ray light curves. 
Flares had been previously observed in \object{GRB~970508} \citep{Piro1999:970508},   
\object{GRB~011121} and \object{GRB 011211} \citep{Piroea2005:011121_011211}. 
\citet{Piroea2005:011121_011211} suggested that the X-ray flares observed in 
the latter two events were due to the onset of the afterglow, since the spectral 
parameters of these flares were consistent with those of their afterglow.
Starting from \object{XRF~050406} \citep{Burrows2005:flarescience,Romano2006:050406}, 
\object{GRB~050502B} \citep{Falcone2006:050502b}, and \object{GRB~050607} 
\citep{Pagani2006:050607}, we have learned that flares can be considerably energetic 
and that they are often characterized by large flux variations. 
Indeed, the flare fluences can be up to 100\% of the prompt fluence and the
flare fluxes, measured with respect to the underlying continuum,  
$\Delta F / F$, can vary in very short timescales $\Delta t / t_{\rm peak}$ 
($\Delta F / F \sim 6$, 500 and 25, $\Delta t / t_{\rm peak} \ll 1$, $\sim 1$, $\sim 1$ 
in XRF~050406, GRB~050502B and GRB~050607, respectively, where $\Delta t$ measures the duration 
of the flare and $t_{\rm peak}$ is measured with respect to the trigger time).
Furthermore, detailed spectral analysis has proven that these flares are spectrally 
harder than the underlying continuum 
\citep{Burrows2005:flarescience,Romano2006:050406,Falcone2006:050502b}. 
In particular, they follow a hard-to-soft evolution, 
which is reminiscent of the prompt emission 
(e.g.\ \citealt{Ford1995:softhard}).
The spectra of the flares in GRB~050502B \citep{Falcone2006:050502b} are better fit by a Band function 
\citep[][which is the standard fitting model for GRB prompt emission]{Band93}, than 
by an absorbed power law (which usually suffices for a standard afterglow). 
Very often multiple flares are observed in the same light curve, with an 
underlying afterglow consistent with having the same slope before and after the flare. 
Finally, \object{GRB~050724} \citep{Barthelmy2005:050724,Campana2006:050724} 
and \object{GRB~050904} \citep{Cusumano2006:050904Nature} 
have demonstrated that flares happen in short GRBs as well as long ones, 
at low and very high redshift (the record being held by GRB~050904 at $z=6.29$). 

The picture that the early detections of flares have drawn was described by 
\citet{flaresprocfull} and \citet[][and references therein]{Chincarini2005:xrt_spain}, 
and a few conclusions were derived, albeit based on a small sample of objects. 
The presence of an underlying continuum consistent with the same slope before and after the flare 
(GRB~050406, GRB~050502B) seems to rule out external shocks models, since no trace 
of an energy injection can be found;
the large observed $\Delta F / F$  cannot be produced by synchrotron self-Compton in the reverse shocks;
the very short timescales $\Delta t / t_{\rm peak} < 1$ also generally rule out external shocks, unless 
very carefully balanced conditions are met \citep[e.g., ][]{Kobayashi2007:IC_reverse};
furthermore, the flare spectral properties (harder than the underlying afterglow, 
evolving from hard to soft) indicate a different physical mechanism from the 
afterglow, and possibly the same as the prompt one.

In this work we present the first comprehensive temporal analysis of all GRBs observed 
by the X-ray Telescope (XRT, \citealt{Burrows2005:XRT})--both long and short, 
independently of whether they are GRBs, X-ray Rich (XRR) or X-ray Flashes 
(XRF, \citealt{Heise2001:XRF_XRR}) 
at low and high redshift--that showed flares in their X-ray light curves. 
We assess whether the evidence for prolonged engine activity accumulated on the first 
observed flares survives statistical investigation and discuss the case that flares 
are indeed related to the workings of the central engine. 
We also present the results of a cross-check analysis between
X-ray flares and pulses detected by the Burst Alert Telescope (BAT, \citealt{Barthelmy2005:BAT}) 
in the gamma-ray prompt emission.
A second paper \citep{Falcone2007:flares_spec} will study the same sample 
from the spectroscopic point of view, in a natural complement to this work. 

This paper is organized as follows. 
In \S\ref{flares:sample} we describe our GRB sample and 
in \S\ref{flares:dataredu} the data reduction procedure; 
in \S\ref{flares:dataanal} we describe our XRT data analysis and in   
\S\ref{flares:batxrt} our cross-check analysis between X-ray flares and pulses 
detected by BAT in the gamma-ray prompt emission.
In \S\ref{flares:results} we present our main results and 
in \S\ref{flares:discussion} we discuss our findings. 
Throughout this paper the quoted uncertainties are given at 90\% confidence level 
for one interesting parameter (i.e., $\Delta \chi^2 =2.71$) unless otherwise stated.
Times are referred to the BAT trigger $T_0$, $t=T-T_0$, unless otherwise specified. 
The decay and spectral indices are parameterized as  
$F(\nu,t) \propto t^{-\alpha} \nu^{-\beta}$, 
where $F_{\nu}$ (erg cm$^{-2}$ s$^{-1}$ Hz$^{-1}$) is the 
monochromatic flux as a function of time $t$ and frequency $\nu$; 
we also use $\Gamma = \beta +1$ as the photon index, 
$N(E) \propto E^{-\Gamma}$ (ph keV$^{-1}$ cm$^{-2}$ s$^{-1}$).

	\section{Sample definition\label{flares:sample}}

We considered all GRBs detected by Swift, INTEGRAL and HETE-2, between the 
Swift launch and 2006 January 31 (119 events) for which XRT obtained a position (99). 
We then examined all light curves searching for large scale activity in 
excess of the underlying power-law light curve (flares).
We defer a detailed analysis of small scale and small frequency variability,
sometimes referred to as ``flickering'' to a later paper. 
None of the INTEGRAL- or HETE-2--triggered bursts showed any flares 
although we note that these bursts were observed by XRT much later than the
Swift-triggered ones. 
As will be discussed in \S\ref{flares:simulations}, where we investigate the
sample biases in depth, we evaluate the completeness of our sample
with a large set of simulations. We established that our flare sample
can be considered complete with respect to faint flares 
only at late times (typically $10^3$ seconds after the trigger). 
In Table~\ref{flares:tab:sample} we list all the GRBs that were 
selected for the analysis along with their redshifts 
(when available, i.e., for 9 of them), $T_{90}$s, and BAT fluences. 
This is what we shall refer to as our ``full'' sample, consisting of
33 GRBs, on which we attempted the timing analysis described 
in \S\ref{flares:dataanal}. 
The light curves of the full sample are shown in Fig.~\ref{flares:bigfig_1a}.

Some light curves, however, were not fit for the full analysis. 
For instance, although joint analysis of BAT and XRT data on GRB~050219A 
\citep{Goad2006:050126_050219A} showed a simultaneous flare 
(hence its inclusion in our sample), the portion of the flare 
that was observed with XRT was not long enough to fully characterize it.
In the same manner, a handful of events (GRB~050826, GRB~051016B, GRB~060109) 
which are included in our full sample because they showed either 
low-signal late-time flares or a flattening in the XRT light curve,  
were excluded from a full analysis because of the low statistics obtained.
All these special cases are reported in Table~\ref{flares:tab:sample} in italics.
After these exclusions, we defined our ``restricted'' sample, which 
consists of 30 GRBs on which we succeeded in performing our full analysis. 

We note that our restricted sample differs from the one of \citet{Falcone2007:flares_spec},
because of different requirements for the analysis. 
As an example, for GRB~050820A \citet{Falcone2007:flares_spec} 
could perform detailed spectroscopic analysis of the flare portion observed by XRT, 
but our full timing analysis was not applicable.

	\section{Data Reduction\label{flares:dataredu}}

The XRT data were first processed by the Swift Data Center at NASA/GSFC into
Level 1 products (event lists).
Then they were further processed with the XRTDAS (v1.7.1) software package,
written by the ASI Science Data Center (ASDC) and distributed within
FTOOLS to produce the final cleaned event lists.
In particular, we ran the task {\tt xrtpipeline} (v0.9.9) applying
calibration and standard filtering and screening criteria.
An on-board event threshold of $\sim$0.2\,keV
was applied to the central pixel of each event, which has been proven
to reduce most of the background due to either
the bright Earth limb or the CCD dark current
(which depends on the CCD temperature).

The GRBs in our sample were observed with different modes, which
were automatically chosen, depending on source count rates, to minimize
pile-up in the data \citep{Hill04:xrtmodes}. For the GRBs observed during
the calibration phase, however, the data were mainly collected in Photon
Counting (PC) mode, and pile-up was often present in the early data.
Furthermore, for a few, especially bright GRBs (which were observed after
the Photo-Diode (PD) mode was discontinued due to a micrometeorite hit on the CCD)
the Windowed Time (WT) data were piled-up, as well.
Generally, WT data were extracted in a rectangular 40$\times$20-pixel region
centered on the GRB (source region), unless pile-up was present.
To account for this effect, the WT data were extracted in a rectangular
40$\times$20-pixel region with a region excluded from its centre.
The size of the exclusion region was determined following
the procedures illustrated in \citet{Romano2006:060124}.
To account for the background, WT events were also extracted within a
rectangular box (40$\times$20 pixels) far from background sources.

The PC data were generally extracted from a circular region with a 30-pixel radius.
Exceptions were made for bright sources, which required a $>30$-pixel radius,
and for faint sources, which required a smaller radius in order to maintain a high
signal-to-noise ratio.
When the PC data suffered from pile-up, we extracted the source events
in an annulus with a 30-pixel outer radius and an inner radius, depending on 
the degree of pile-up as determined via the PSF
fitting method illustrated in \citet{Vaughan2006:050315}.
PC background data were also extracted in a source-free circular region.

For our analysis we selected XRT grades 0--12 and 0--2 for PC and WT data,
respectively (according to Swift nomenclature; \citealt{Burrows2005:XRT}).
To calculate the PSF losses, ancillary response files were generated with 
the task {\tt xrtmkarf} within FTOOLS, and account for different extraction 
regions and PSF corrections. We used the latest spectral redistribution matrices 
in the Calibration Database maintained by HEASARC. 

From both WT and PC data, light curves were created in the 0.2--10\,keV energy band 
using a criterion of a minimum of 20 source counts per bin, 
and a dynamical subtraction of the background. Therefore, in our sample, 
each light curve was background-subtracted, and corrected for pile-up, vignetting, 
exposure, and PSF losses.

	\section{Data Analysis\label{flares:dataanal}}
The first goal of this work was to obtain a quantitative assessment of flare characteristics.
We thus set to measure statistical parameters such as 
the ratio of the flare duration to the time of occurrence $\Delta t/t$, 
the power-law decay slope $\alpha_{\rm fall}$, 
the decay to rise ratio $\Delta t_{\rm fall}/\Delta t_{\rm rise}$, 
the flare energetics, and the flare to burst flux ratio. 
Different approaches suited the data best, depending on the flare statistics, 
as we outline below.

\subsection{Equivalent widths\label{flares:ew}}
%

We calculated the equivalent width of the flares defined as
$EW = \int \frac{F_{\rm observed}(t) - F_{\rm continuum}(t)}{F_{\rm continuum}(t)}\,dt$,
where $F_{\rm continuum}(t)$ describes the assumed shape of the continuum  
light curve underneath the flare (the local power law ``underlying continuum'') 
and $F_{\rm observed}(t)$ is the observed light curve,
i.e., the combination of the continuum and flare
(the analytical fits to the continua are described in detail in \S\ref{flares:gaussian_fits}
and their parameters reported in Table~\ref{flares:tab:cont_pars}).
The equivalent width 
(expressed in units of seconds, as reported in Table~\ref{flares:tab:fits} column~6) 
represents the time needed for integration of the continuum to collect the same fluence as 
of the flare and it can give us a first indication of the lowest fluence we are able to measure 
for a flare. 
Indeed, the faintest equivalent width measured, on a rather weak afterglow with 
XRT fluence of $\sim 1.3\times10^{-8}$ erg cm$^{-2}$ light curve, is 
7.9\,s in a small flare detected in GRB~050819.  
At the other extreme of the $EW$s is GRB~050502B, where we detect two flares, 
both characterized by large $EW$s.
The first one is extremely bright and indeed has a fluence that is larger than the 
fluence of the underlying continuum light curve 
($1.43\times10^{-6}$ erg cm$^{-2}$ and $1.23\times 10^{-6}$ erg cm$^{-2}$,  respectively). 
Even though (see \S\ref{flares:simulations}) our completeness for faint flares is somewhat 
limited at early times, this may be an indication that the flare is generally stronger 
than the continuum light curve and possibly an unrelated phenomenon. 

Our ability to measure $EW$s is limited by the discrete sampling of the light curves as well as
the relative faintness of the flares, therefore we could only obtain $EW$s for 
48 flares. Figure~\ref{flares:fig:ew_distrib} shows the distribution of the $EW$s
for our sample. 

\subsection{$\Delta t/t$ from Gaussian fits\label{flares:gaussian_fits}}

The simplest analytical characterization of the flare morphology is obtained by 
adopting a multiply-broken power law to model the underlying continuum, 
and a number of Gaussians to model the superimposed flares. 
We adopted the following laws for the continuum,  
{\it i)} simple power law: $F(t) = K t^{-\alpha_1}$, 
{\it ii)} broken power law: $F(t) = K t^{-\alpha_1}$ for $t<t_{\rm b1}$ and  
$F(t) = K\,t_{\rm b1}^{-\alpha_1} \, (t/t_{\rm b1})^{-\alpha_2}$ for $t>t_{\rm b1}$,  
{\it iii)} doubly-broken: $F(t) = K t^{-\alpha_1}$ for $t<t_{\rm b1}$ and  
$F(t) = K\,t_{\rm b1}^{-\alpha_1} \, (t/t_{\rm b1})^{-\alpha_2}$ for $t_{\rm b1}<t<t_{\rm b2}$, 
 $F(t) = K\,t_{\rm b1}^{-\alpha_1} \, (t_{\rm b2}/t_{\rm b1})^{-\alpha_2} \, (t/t_{\rm b2})^{-\alpha_3}$ 
for $t>t_{\rm b2}$, and so on, where $t_{\rm b1}$ and  $t_{\rm b2}$ are the times of the breaks. 
For our flares, we iteratively added as many Gaussians as required 
to accommodate the $\chi^2$ locally around each flare.
The best-fit model parameters for each component (continuum and flares) 
were derived with a joint fit and are reported in Table~\ref{flares:tab:cont_pars} 
(continuum parameters) and Table~\ref{flares:tab:fits} (flare parameters, columns~2--4).
Column~5 of Table~\ref{flares:tab:fits} reports flare peak fluxes measured with respect to 
the underlying continuum, or $\Delta F/F$). 
The full gallery of fits is illustrated in Fig.~\ref{flares:bigfig_1a}. 
In Fig.~\ref{flares:fig:peakt_distrib} we show the distribution of the peak times 
(i.e., the Gaussian peaks). 

Based on these fits, we calculated $\Delta t/t$ for each flare, 
adopting the Gaussian width ($\sigma$) and peak $t_{\rm peak}$ as $\Delta t$ and $t$, respectively,
where $t_{\rm peak}$ ranges between 95\,s and $\sim 75$\,ks. 
We do not include the Gaussian fits for GRB 060124 for sake of homogeneity of the sample, 
since the XRT data include the prompt phase (\citealt{Romano2006:060124}).
Our ability to fit flares with Gaussians is less affected by discrete sampling of the light 
curves than for the $EW$ determination, but it still suffers from the faintness of the flares;
therefore we obtained fits for 69 Gaussian-modeled flares. 
In Fig.~\ref{flares:fig:dt_t_gauss_distrib}, we show the distribution of 
the $\Delta t/t$, which peaks at $0.13$, 
and which yields a mean value of $\langle \Delta t/t \rangle = 0.13\pm 0.10$. 
An assessment of selection effect that may affect this result is reported in 
\S\ref{flares:simulations}.

\subsection{Decay slopes, rising and decaying times from more realistic models\label{flares:alltaus}}

Flare profiles can be quite complex. As an example, in Fig.~\ref{flares:morphology4} 
we show the light curves of GRB~050730, in which different flares are
best fit by different laws (two power laws for the first, and an exponential rise 
followed by a power-law decay for the second one), of  GRB~050502B (first flare), 
and GRB~060111A. 
A more realistic description of the flare profile should therefore account 
for the skewness observed in many flares as well as different rising and 
falling slopes and times, which we indicate with 
$\alpha_{\rm rise}$, $\alpha_{\rm fall}$, $\Delta t_{\rm rise}$, and $\Delta t_{\rm fall}$, respectively. 
Such a fit can be performed with power laws ($F(t)\propto (t-t_{0})^{-\alpha}$), 
in which case it is critical to define the reference time $t_0$.
In practice, for $t_0$ we consider the peak time as well as the times, 
before and after the peak, when the flare profile deviates significantly from 
the continuum fit: $\Delta t_{\rm rise}=t_{\rm peak} - t_1$, 
$\Delta t_{\rm fall}=t_2 - t_{\rm peak}$, 
where $t_1$ and $t_2$ are the times when the flare emission (fitted in
this case by a Gaussian) crosses the fraction $f$ of the flare peak emission 
on either side of the peak. 
For the calculation of the decay slopes, we chose $f=0.01$. 
As for the previous fits (\ref{flares:ew}, \ref{flares:gaussian_fits}), 
our procedure requires a power law continuum beneath each flare. 
The values of $\alpha_{\rm fall}$ we computed for this sample (consisting of 35 flares) 
are reported in Table~\ref{flares:tab:fits} (column~7) and their distribution is
shown in Fig.~\ref{flares:fig:slope_distrib}.
We derive $\langle \alpha_{\rm fall} \rangle =3.54$ with standard deviation of $\sigma=1.50$. 
We note that our choice of $f=0.01$ was an operative decision; 
using a different definition for $t_1$ and $t_2$, the measure of the slope decay changes. 
For instance, for the large flare in GRB~050502B we obtain $\alpha_{\rm fall}=6.32$, 5.58,
and 5.20 for $f=0.01$, 0.05 and 0.10, respectively.
This shows how critical the definition of  $t_1$ and $t_2$ 
is in measuring the  decay slope that describes the temporal behaviour of a burst or flare.

The quantities $\Delta t_{\rm rise}$ and $\Delta t_{\rm fall}$ are in themselves 
quite interesting, since, as is well known from the work of \citet{Norris1996:pulses} and
from the simulations by \citet{Daigne1998:internal_shocks}, 
the observed bursts, which are due to the internal shocks, present a Fast-Rise 
Exponential-Decay (FRED) shape with 
a ratio $\Delta t_{\rm fall}/\Delta t_{\rm rise}=3.4$.  
We calculated $\Delta t_{\rm rise}$ and $\Delta t_{\rm fall}$ for our sample 
defining them in terms of $f=0.05$, by assuming the underlying power law 
continuum beneath the flares, and performing a separate fit to the rising and 
decaying part of the flare light curve. 
While for the decaying part we always used a power law, we found that 
in many instances the best fit to the rising part was obtained with an exponential. 
Using these fits, we calculated $\tau_{90}$ (the time defined by $f=0.05$)
and the ratio $\Delta t / t$ adopting $\Delta t=\tau_{90}$ and $t=t_{\rm peak}$.
Table~\ref{flares:tab:fits} reports $\tau_{90}$, $\Delta t_{\rm fall}/\Delta t_{\rm rise}$,
$\Delta t/t$ (columns~8--10), while Fig.~\ref{flares:fig:rise_fall_distrib} shows the 
distributions of $\Delta t_{\rm fall}/\Delta t_{\rm rise}$.

\subsection{Selection effects\label{flares:simulations}}
As stated above, our ability to measure statistical quantities from the light curves 
critically depends on both the discrete sampling of the light curves as well as
the actual intensity of the flares with respect to the continuum beneath them. 
In this section we present our considerations on the biases 
that may affect our analysis and their effect on our results. 
One of the first difficulties comes from the blending of flares,
which causes the $EW$, $\Delta t$, $\Delta t /t$, to be overestimated. 
Our result of low $\Delta t /t$ is thus an upper limit on the 
intrinsic sharpness of flares.

\subsubsection{Time resolution and low-earth orbit biases\label{flares:bias_time}}

The time resolution of our observations, which decreases logarithmically 
during the XRT afterglow follow-up, is the first critical factor. 
Typically, at the beginning of the XRT light curve the sampling is quite good, 
but if the flare duration is of the order of the time it takes it to fade,
then it will not be possible to recognize it as such, and it will be interpreted 
as a steep power law, instead. 
This was often observed in the early XRT light curves, as reported by 
\citet{Tagliaferri2005:nature} and \citet{Obrien2006:xrtbat}, and it is 
partially related to the short but finite time (usually $>60$\,s) it takes 
Swift to re-point to the GRB. 
On the other hand, at the end of the XRT light curve, the sampling also degrades
because of the long integration required to achieve sufficient $S/N$, 
so that flares shorter than the integration
time are smeared out and consequently, except for the brightest ones,
their resulting average count rate drops below the detection threshold.

Due to Swift's low-earth orbit, the data are not collected in a continuous way,
but in portions of an orbit that last less than an hour. 
This is illustrated in Fig.~\ref{flares:fig:bintimes} (left), which represents the 
distribution of the observing times relative to the BAT trigger, of all the light curves 
in our sample. 
For each observation of the light curve, we estimated the time, which we shall refer to as
bin time (BT), within which the counts were accumulated in order to have a $S/N>3$.
For $t>10^4$\,s the BT will generally include data from consecutive orbits.  
In Fig.~\ref{flares:fig:bintimes} (right), we show the time resolution (BT) as a function of 
the time since the BAT trigger, as well as the curve that 
corresponds to BT$/t=0.1$ and lies above the large majority of the data.
It indicates that the instrumental resolution $BT/t=(\Delta t / t)_{S/N=3}$ 
is in most cases significantly better than $\Delta t/t\approx0.1$ and is often 
even better than 0.01. 
In other words, our data are not biased against $\Delta t/t\la0.1$.

\subsubsection{Biases in the sample definition criteria\label{flares:bias_sample}}

In order to evaluate the completeness of our sample  
we tested the sample definition criteria 
against selection effects by means of simulations.

First of all, for each flare in our sample, we evaluated the signal to noise (S/N)
ratio as the ratio between the fluence of the flare and the continuum
calculated in the time interval [$-1\sigma$,$+1\sigma$], where $\sigma$ is the Gaussian
width. 
The minimum detected S/N is 5. 
Then, to simulate our procedure, we first calculated the median continuum light curve 
from the whole data sample. This
median light curve at late times is well described by a single power law with
$\alpha_{\rm median}=1.1$.  On top of that we summed a Gaussian flare
with the 3 parameters randomly chosen and uniformly distributed 
over large intervals which
fully contain the real data parameter values.  From this parent
distribution we generated a collection of photons. Finally we
reconstructed the light curve, using the same procedure as we used for
the real data. In this way we realistically reproduced a typical
observed light curve.  The only significant difference is that when we
simulated, we assumed a continuous observation, whereas the real
observation is split in different orbits. However, as discussed in the
previous section, this assumption does not affect our conclusions. We
repeated the test 14000 times in order to have a statistically
significant sample of simulated light curves.  For each randomly
generated peak we calculated the S/N ratio and we flagged it as
identified when its S/N ratio exceeded the value of 5 and, at the same
time, at least 3 points in the light curve lay more than 2-$\sigma$ above the 
continuum.  In Fig.~\ref{flares:fig:pianosimule} we plot the
results of the simulations in the ($t$, $\Delta t$) plane. 
For each ($t$, $\Delta t$) value 
we could assign a detection probability;
the points are the real data. We note that at early times our
sample data lie in the region with low detection probability. This is
clearly an effect of the significance threshold: at the beginning, the
afterglow is brighter, the absolute level of the noise is high and a
flare can be detected only if it is bright enough to have significance
above the threshold.  Given the median continuum light curve, our 
simulations show that 
 if flare has a $\sim$90\% detection probability at 10 ks, 
at 300s it will have a $\sim$ 30\% detection probability.  
At late times the simulation results show that
the detection probability decreases with smaller $\Delta t$
(bottom-right corner of the plot): this is an effect of threshold  
set as the minimum number of photons per bin of the light curve.
At larger times the light curve has a sparse sampling
and a faint and narrow flare produces only few bins over the
continuum. Our simulations also show that, in the region of the
plane defined by $\Delta t/t > 2\times 10^{-3}$ and by 
$t>10^4$\,s, the detection probability is uniformly larger than 90\%.
In Fig.~\ref{flares:fig:pianosimule} we also plotted the line
$\Delta t=t$ over which we do not expect to find any flare.  Comparing
our sample with the simulation probability map we conclude that we do
not find narrow flares at large $t$ in the areas of the parameter plane
where we have very high detection probability. 
Therefore, although we cannot evaluate the completeness of our sample at
early times, from our simulations we can firmly conclude that 
the lack of narrow flares at late times (typically $10^{3}$ s) 
is not due to incompleteness.

\section{XRT flares vs. BAT pulses\label{flares:batxrt}}

We investigated if there is a clear link between the properties
of the pulses detected in the gamma-ray burst profile
by BAT in the 15--350 keV band and the X-ray flares as detected by
XRT.

In order to define a procedure to select and characterize BAT
pulses, we used an adapted version of the criterion defined by
\citet{Li1996:lognormal_distributions}:
we started from the
64-ms mask-tagged light curve extracted following the standard
BAT pipeline and searched for those bins whose count rates
exceed $m$ contiguous bins by $n\sigma$ on both sides.
We applied this procedure with three different combinations
of $(m,n)$: $(5,3)$, $(3,4)$, and $(1,5)$ and to all of the curves
with multiple binning times from 64 ms to 32 s, taking into
account all of the possible shifts at a given binning time.
This choice proved to be effective in catching different pulses
clearly detected by visual inspection.
We assessed the false positive rate of pulses so detected
with a Monte Carlo test: we took the number of 64-ms bins of the
longest GRB light curve available and simulated 100 synthetic
light curves with constant signal, whose count rates were affected
by Gaussian noise. We applied the same procedure to these 100
synthetic light curves and found 8 false pulses.
We then estimated the average false positive rate as of 0.08
fake pulses for each GRB light curve.
As we collected 28 GRBs with a complete BAT light curve [GRB~050820A
was ignored because Swift entered the South Atlantic Anomaly (SAA) before
the gamma-ray prompt emission ceased], we expect about 2 false pulses.
We detected 46 pulses distributed in 28
gamma-ray profiles, so we can safely assume a negligible
contamination of the gamma-ray pulses sample due to statistical fluctuations.

Table~\ref{flares:tab:bat_peaks} shows the results of the BAT
pulses quest, which identified 46 pulses out of 28 GRBs.
For each pulse, columns 1--6 report: (1) the GRB name it belongs to,
(2) the ordinal number of the pulse within the GRB, (3) the binning time
used to detect the pulse (which also corresponds to the
uncertainty on the peak time), (4) the peak time, (5) the peak rate
(counts s$^{-1}$),
(6) error on the peak rate (counts s$^{-1}$).

We do not find any clear correlation between the number of gamma-ray
pulses and the number of X-ray flares.
Column (1) in Table~\ref{flares:bat_xrt_pulses} reports
the number of gamma-ray
pulses found in a given burst; column (2) the number of
X-ray flares, and column (3) reports the number of GRBs
with that combination of numbers of pulses and flares.
The most common case is when the burst exhibits one single
pulse followed by one, or two X-ray flares.

We tested whether there is any statistical evidence that GRBs with
many/few pulses are more likely to have many/few X-ray flares.
Let $n_{\gamma}$ and $n_{\rm x}$ be the number of gamma-ray pulses
and of X-ray flares of a given burst, respectively.
We split the sample in two classes in two ways:
those with $n_\gamma\ge2$ (``many pulses''; 10 GRBs) and
those with $n_\gamma<2$ (``few pulses''; 18 GRBs);
likewise, those with $n_{\rm x}\ge3$ (``many flares''; 11 GRBs)
and those with $n_{\rm x}<3$ (``few flares''; 17 GRBs).
From Table~\ref{flares:bat_xrt_pulses} one counts 5 bursts
with both many pulses and many flares. In the assumption of no
correlation between the number of pulses and the number of
flares, the probability of choosing randomly $n\ge5$ bursts with
many pulses out of 11 bursts with many flares is about 35\%: i.e.
given a burst with many flares, nothing can be inferred about
its number of pulses.
Similarly, the probability of selecting $n\ge5$ bursts with
many flares out of 10 bursts with many pulses is 32\%: i.e.
given a burst with many pulses, nothing can be inferred about
its number of flares.
We also tried to split the sample with different combinations
of thresholds on $n_\gamma$ and $n_{\rm x}$, but no statistically
significant correlation has come out.
Furthermore, we compared the distributions of the numbers of
pulses derived for the two populations, i.e. those with few flares
and those with many flares. A Kolmogorov-Smirnov (KS) test shows
no difference between the two subsets, with 88\% probability
that they have been drawn from the same population.
Likewise, we compared the distributions of the numbers of
flares derived from splitting the sample between GRBs with
few and many pulses, respectively.
According to the KS test, we cannot reject that the two distributions are
the same at 99\% confidence level.
We conclude that one cannot infer anything about the number
of X-ray flares from the number of gamma-ray pulses and
vice versa.

We also compared the distribution of the numbers of pulses
with that of the numbers of flares and a KS test does not
prove any significantly different origin (30\% probability
of having been drawn from the same distribution).

We also sought any possible correlation between the intensity
of the pulses and properties of the flares as well as between the peak times
of either class.
To this aim, for each GRB in Table~\ref{flares:tab:xrtbat_times} we grouped 
the following pieces of information: columns 1--3 report the GRB name,
the number of BAT pulses $n_{\gamma}$ and the number of X-ray flares
$n_{\rm x}$, respectively.
From column (4) up to column (12) the correspondent times are reported
(referred to the BAT trigger time): the first $n_{\gamma}$ refer to
the BAT pulses, while the remaining $n_{\rm x}$ refer to the X-ray flares.

For either class we considered those bursts with at least two events
(i.e., either two pulses or two flares).
We searched for any correlation between the
quiescent time (between two successive pulses, or between two flares)
and the peak brightness of the following event, but our search was unsuccessful.
We also studied the relation between quiescent time and the ratio
of the following peak, ${\rm peak}_{i+1}$ over the preceding
peak, ${\rm peak}_i$. Figure~\ref{fig:quiesc_vs_ratio} shows
two interesting results: firstly, there is no clear dependence of
this ratio on the quiescent time for both classes.
Secondly, the distribution of ratios derived from the X-ray flares
is consistent with that of the gamma-ray pulses. In particular,
if we merge the two sets of ratios, this is consistent with a log-normal
distribution with mean value $<\log{({\rm peak}_{i+1}/{\rm
peak}_{i})}>=-0.258$
and $\sigma_{\rm log}=0.68$ (see Fig.~\ref{fig:logallratio}).
If we ignore the two points due to X-ray flares
with the lowest ratio (see Fig.~\ref{fig:quiesc_vs_ratio}), the mean value
and standard deviation turn out to be $-0.157$ and 0.41, respectively (shown
in Fig.~\ref{fig:quiesc_vs_ratio}).

We therefore conclude that the relation between successive
pulses and between successive flares is the same: in particular, on
average the next event has a peak $10^{-0.157}\simeq0.7$ times as high
as the preceding, while the scatter is between 0.3 and 1.8.
This further piece of evidence points to a common origin for gamma-ray
pulses and X-ray flares.

\section{Results \label{flares:results}}

In this section we explore possible correlations between the parameters 
derived in the analysis and summarize our findings. 

\subsection{Gaussian peak time--intensity correlation\label{flares:tpeak_norm_correlation}}

We tested for a correlation between the Gaussian peak intensity and the peak position 
(s since the BAT trigger).
As shown in Fig.~\ref{flares:fig:peakt_norm_corr}, the correlation is 
strong, with a Spearman rank coefficient $r_{\rm s}= -0.539$ (number of points 
$N=63$, and null hypothesis probability nhp$=5.24\times10^{-6}$). 
However, it can be argued that this correlation is driven by the flares at late times 
and that there is large scatter for $t<10^3$\,s. 
In this light, this would be an indication that the mechanism producing the flares 
holds no memory of when the trigger time occurred. 
Therefore, the only firm conclusion we can draw is that 
the late flares have a peak intensity which is less than the early ones and
coupling this with the $\Delta t$ results (see \S\ref{flares:gaussian_fits}) 
we infer that late flares have a lower peak intensity but last  much longer so their fluence can be
very large.

\subsection{$EW$  correlations\label{flares:ew_tpeak_correlation}}

We find a strong correlation between the equivalent width and the time 
of the occurrence of the flare, $t_{\rm peak}$,   ($r_{\rm s} = 0.729$, $N=48$, nhp$=4.1\times10^{-9}$)
which is mostly due to the large dynamical range in $t_{\rm peak}$ values. 
Indeed, we find no correlation of $EW/t_{\rm peak}$ with $t_{\rm peak}$. 
There is also no correlation between $EW/t_{\rm peak}$ and $\Delta t /t_{\rm peak}$ 
(Fig.~\ref{flares:EW_Peak_Time}) which is probably a further indication that the flares 
are not related to the underlying continuum and that they originate from the
engine rather than the external shock.
We also note that $EW/t_{\rm peak}$ is generally greater or equal to $\Delta t /t_{\rm peak}$ 
(solid line in Fig.~\ref{flares:EW_Peak_Time})
because the $EW$ calculation is sensitivity-limited. 
The median value of $EW/t_{\rm peak}$ is 0.5 (mean value 5.7 with standard deviation 25.5).

\subsection{Decay slope-time correlation\label{flares:decayslope_t_correlation}}
%
If we consider  $\alpha_{\rm fall}$  as a function of time, we obtain, for 
$t <10000$\,s, that $\alpha_{\rm fall}=2.45+0.418\,t$. 
The correlation is only marginal  ($r_{\rm s} = 0.152$, $N=35$, nhp$=0.382$) 
and a somewhat smaller value is 
obtained by using, as stated above, $f=0.05$.
We conclude therefore that in most cases the exponent of the power law decay is
in agreement with the curvature effect \citep{Kumar2000b:naked_grbs}. 
In late internal shock models, $T_0$ has to be reset every time when the 
central engine restarts \citep{Zhang2006:theory_from_xrt}.
As shown in \citet{Liang2006:curvature}, if one assumes that the post-flare
decay index satisfies the curvature effect prediction $\alpha=\beta+2$, the
required $T_0$ is right before the corresponding X-ray flares at least for
some flares. This lends support to the curvature effect interpretation and
the internal origin of the flares.
In a few flares, however, the
giant flare observed in GRB~050502B being the best example, 
the decay slope is much steeper if $T_0$ is put near the peak 
(see, \citealt{Dermer2004:curvature,Liang2006:curvature}).

\subsection{$\Delta t_{\rm fall}/\Delta t_{\rm rise}$--$\tau_{90}$ 
	correlation\label{flares:tf/tr_t90_correlation}}

During the prompt emission, as tested by \citet{Norris1996:pulses}, 
shorter bursts tend to be more symmetric and the width of the burst tends to be correlated with 
$\Delta t_{\rm fall}/\Delta t_{\rm rise}$ in the sense that longer bursts tend to show 
a larger ratio, or $\Delta t_{\rm fall}/\Delta t_{\rm rise}\sim 2$--3, a value that agrees 
quite well with the mean  $\langle \Delta t_{\rm fall}/\Delta t_{\rm rise} \rangle =2.35$.
This effect has been quite clearly simulated by \citet{Daigne1998:internal_shocks}. 
The flare sample was used to test for this effect.
We used $\tau_{90}$ as a reference time to minimize the bias 
we may have in the curve subtraction when the signal of the flare is weak. 
In addition, we considered both expressions such as $F(t)=a (1-bx^{-c})$ 
and simple power laws ($F(t)=k~x^{-m}$) to model the sides of the flares.
Using $f=0.05$ the difference in the width ($\tau_{90}$) of the flare
inferred from the two fits is negligible for the scope of this work. 

As shown in Fig.~\ref{flares:taus_t90} we find a tentative correlation 
between the ratio
$\Delta t_{\rm fall}/\Delta t_{\rm rise}$ and  $\tau_{90}$
($r_{\rm s} = 0.543$, $N=24$, null hypothesis probability, nhp$=6.15\times10^{-3}$). 
Such a correlation was pointed out by  
\citet{Daigne1998:internal_shocks} in their simulations of the prompt emission.

\subsection{Summary of Results\label{flares:summ_res}}

We gathered a sample of 33 light curves drawn from all 
GRBs detected by Swift, INTEGRAL and HETE-2, which had an XRT follow-up 
and which showed either large-scale flaring or small scale 
(mini-flaring) flickering activity. 
None of the INTEGRAL- or HETE 2-triggered bursts showed any flares
(however, note that these burst were observed by XRT much later than the
Swift-triggered ones). 
For 30 of these bursts, we performed a full
statistical analysis, by fitting the continuum light curve beneath the flares 
(the XRT canonical light curve shape)
with a multiply-broken power law and the flares with a sample of analytical functions. 
Our sample of Gaussian fits consists of 69 flares, for 48 of which we calculated $EW$s 
by numerical integration, for 35 we could determine a decay slope, and 
for 24 of them $\tau_{90}$, $\Delta t_{\rm fall} / \Delta t_{\rm rise}$ and $\Delta t / t$.  
Our results can be summarized as follows.

\begin{enumerate}
\item 	Flares come in all sizes and shapes and can be modelled with Gaussians 
	(symmetrically shaped) superimposed on a multiply-broken power-law  underlying continuum. 
	However, for a more accurate description, in many instances 
	an exponential rise followed by a power law decay 
	or power law rise followed by a power law decay is required to produce good fits. 

\item Flares are observed in all kinds of GRBs: 
	long (32 GRBs) and short (2 GRBs),
	high-energy--peaked or XRFs  (32 vs.\ 2);
	they are found both in early and in late XRT light curves. 

\item The equivalent widths of our sample, which measure the flare fluence in terms of the underlying continuum,  
	range between 8\,s  and $7\times 10^{5}$\,s. 

\item The distribution of the ratio $\Delta t / t$,  as defined by the width and peak of the Gaussians flare models, 
	yields $\langle \Delta t / t \rangle = 0.13\pm0.10$. 
	Our simulations show that our time resolution allows us to sample flares that may have  
	$\Delta t / t <0.1$, so that the above values are not the result of the biases in our sample 
	or our fitting procedures. 
	Our simulations also show that there are no sharp (small $\Delta t / t$) flares at large times. 
\item The decay slopes $\alpha_{\rm fall}$ range between 1.3 and 6.8 and generally agree with the curvature effect.  
\item The ratio of decay and rise times range between 0.5 and 8.
\item Correlations are found between 
	\begin{enumerate}
	\item $t_{\rm peak}$--peak intensity (strong); 
	\item $EW$--$t_{\rm peak}$ (very strong);
	\item $\alpha_{\rm fall}$--$t$ (poor);
	\item $\Delta t_{\rm fall}/\Delta t_{\rm rise}$--$\tau_{90}$ (tentative).
	\end{enumerate}

\item We do not find any clear correlation between the number of gamma-ray
	pulses and the number of X-ray flares. One cannot infer anything about the number
	of X-ray flares from the number of gamma-ray pulses and vice versa.
	We also conclude that the relation between successive
	pulses and between successive flares is the same: in particular, on
	average the next event has a peak $10^{-0.157}\simeq0.7$ times as high
	as the preceding, while the scatter is between 0.3 and 1.8.
	This is a piece of evidence pointing to a common origin for gamma-ray
	pulses and X-ray flares.

\end{enumerate}

\section{Discussion  \label{flares:discussion}}

The analysis of the flares in the present sample together with the 
revisiting of the canonical XRT light curve 
\citep{Chincarini2005:lcvs,Nousek2006:lcvs,Obrien2006:xrtbat,Zhang2006:theory_from_xrt}
make it clear that the onset of the XRT observation corresponds to 
the late tail of the prompt emission as defined in the current model. 
Flares are often observed in the early XRT light curves. 
Their slopes do not conflict with the curvature effect limit; they simply need a 
different interpretation and a proper location of $T_0$ \citep{Liang2006:curvature}.

A similar reasoning explains the decay slope of the flares. 
We have seen, in agreement with the finding of \citet{Liang2006:curvature}, 
that the decay slope is very sensitive to the definition of $T_0$ and that if this is 
located at the beginning of the flares we are within the constraint of 
the curvature effect. This essentially means that the shock, after reaching the maximum 
luminosity, is not fed anymore and fades out. 
Some of the uncertain or critical cases of flares may be due to the presence of blends. 
Blends and superimposed mini flares are indeed very common and we can observe 
them very clearly in all those cases in which the statistics are very good. 
Although the analysis may be affected in part by this contamination, the results remain robust. 
Indeed, the contamination makes our results even more robust since the detection 
of unseen blends would make the selected $T_0$ large, thus decreasing the 
measured slope and width of the flares. 

We also considered the possibility of a correlation between the characteristics 
of the prompt emission as observed by BAT and the frequency of flares detected by XRT. 
We found no correlation. This simply means that the flares are random events and are 
not related to the way the prompt emission develops in time. 
For instance, there could be an initial flickering, due to the collision 
of highly relativistic shells followed by random flare events due to the collision of 
slower residual pellets, as discussed below. The contamination to our sample 
due to the fact that some of the early XRT flares are the tail of the late 
prompt emission does not change this result. However, this 
needs to be further investigated using a larger statistical sample. 

Furthermore, we have shown that our analysis is not affected by bias in the detection 
of high-intensity late flares and that such flares never show a peak of intensity 
as strong as those observed in the early flares. 
On the other hand, due to their rather long duration, these flares are also 
very energetic.

Most of the indications we have so far seem to lead toward an activity that 
is very similar to that of the prompt emission, with flares that are 
superimposed on a very standard light curve. This has been observed both in long and short bursts. 

In light of the calculations of \citet{Ioka2005:flare_diagnostic}, we calculated $\Delta F / F$
and $\Delta t / t$ 
values from our flare sample and plotted them over the 
kinematically allowed regions for afterglow variabilities, as shown in Fig.~\ref{flares:yoka}. 
\citet{Ioka2005:flare_diagnostic} distinguish between four cases: 
(a) dips, arising from non-uniformity on the emitting surface induced, e.g., by density fluctuations, 
[eq.\ (4) in \citet{Ioka2005:flare_diagnostic}];
(b) bumps due to density fluctuations 
	\citep{Wang2000:variability_turbulence,Lazzati2002:density_waves,Dai2002} 
	[eq.\ (7) in \citet{Ioka2005:flare_diagnostic}];
(c) bumps due to patchy shells \citep{Meszaros1998,Kumar2000:grb_LF}, for which $\Delta t >t$; 
(d) bumps due to refreshed shocks 
	\citep{Rees1998:refreshed_shocks,Panaitescu1998:refreshed_shocks,Kumar2000:shock_models}, 
for which $\Delta t > t/4$. 

Our findings are consistent with the conclusion  
of \citet{Zhang2006:theory_from_xrt} and \citet{Lazzati2007:flares}, the latter 
based on a preliminary presentation of our dataset in  \citet{Chincarini2006:vulcano}, 
i.e.,  a sizable fraction of the flares cannot be related to external shock mechanisms.
In particular, only one point (corresponding to a flare in GRB~051117A) lies in the region 
of  $\Delta t > t$, where flares are consistent with the patchy shells model;
only three points (including the the early flare of GRB~050502B)
lie in the region of flares that can be caused by ambient density fluctuations; 
finally, only 29/69 lie in the region that describes flares due to refreshed shocks. 
Among the rest, 10/69 can only be due to internal shocks. 

\citet{Perna2006} proposed that X-ray flares are due to accretion of a
fragmented disk. Due to viscous evolution, blobs far from the central
black hole takes longer to be accreted and are therefore more spread-out
when accretion occurs. The accretion rate is correspondingly lower. This
naturally gives a peak luminosity-flare epoch anti-correlation as has
been revealed by the data. This same merit could be retained if a magnetic
barrier modulate a continuous accretion flow near the black hole at
different epochs \citep{Proga2006}.

\acknowledgments

This work is supported at OAB by ASI grant I/R/039/04 and PRIN 2005025417, 
at Penn State by NASA contract NAS5-00136, 
and at the University of Leicester by PPARC. 
We gratefully acknowledge the contributions of dozens of members 
of the XRT and BAT teams at
OAB, PSU, UL, GSFC, ASDC, and MSSL and our subcontractors, 
who helped make these instruments possible.

{\it Facilities:} \facility{Swift (XRT)}.


\clearpage

\begin{deluxetable}{llccccr} 
  \tablewidth{0pc} 	      	
  \tablecaption{GRB XRT light curve sample.\label{flares:tab:sample}} 
  \tablehead{	
\colhead{GRB\tablenotemark{a}} & \colhead{Redshift} & \colhead{$T_{90}$} & \colhead{BAT Fluence\tablenotemark{b}} & \colhead{Reference} & \colhead{Reference} & \colhead{Notes} \\
\colhead{Name} & \colhead{} & 	\colhead{(s)} & \colhead{(erg cm$^{-2}$)} & \colhead{redshift} & 	\colhead{BAT}  & \colhead{} \\
\colhead{(1)} 	 & \colhead{(2)}  & \colhead{(3)}  & \colhead{(4)}  & \colhead{(5)}  & \colhead{(6)}  & \colhead{(7)} 
}
  \startdata	
050406	&	\nodata	&	$5\pm1$		&	$9.0\times10^{-8}$		&	\nodata	&	1	& XRF	\\
050421	&	\nodata	&	$10.3\pm2$	&	$(1.8\pm0.7)\times10^{-7}$	&	\nodata	&	2	&	\\
050502B	&	\nodata	&	$17.5\pm0.2$	&	$(8.0\pm1.0)\times10^{-7}$	&	\nodata	&	3	&	\\
050607	&	\nodata	&	26.5		&	$(8.9\pm1.2)\times10^{-7}$	&	\nodata	&	4	&	\\
050712	&	\nodata	&	$48\pm2$	&	$1.8\times10^{-6}$		&	\nodata	&	5       &	\\
050713A	&	\nodata	&	$70\pm10$	&	$(9.1\pm0.6)\times10^{-6}$	&	\nodata	&	6	&	\\
050714B	&	\nodata	&	55.		&	$(6.5\pm1.4)\times10^{-7}$	&	\nodata	&	7	& XRF	\\
050716	&	\nodata	&	$69\pm1$	&	$(8.3\pm1.3)\times10^{-6}$	&	\nodata	&	8	&	\\
050724	&	0.258	&	$3\pm1$		&	$(6.3\pm1.0)\times10^{-7}$	&	9	&	10	& Short	\\
050726	&	\nodata	&	30.		&	$(4.3\pm0.7)\times10^{-6}$	&	\nodata	&	11	&	\\
050730	&	3.967	&	$155\pm20$	&	$(4.4\pm0.4)\times10^{-6}$	&	12	&	13	&	\\
050801	&	\nodata	&	$20\pm3$	&	$(4.4\pm1.0)\times10^{-7}$	&	\nodata	&	14	&	\\
050802	&	\nodata	&	$13\pm2$	&	$(2.8\pm0.3)\times10^{-6}$	&	\nodata	&	15	&	\\
050803	&	0.422	&	$85\pm10$	&	$(3.9\pm0.3)\times10^{-6}$	&	16	&	17	&	\\
050814	&	\nodata	&	$65^{+40}_{-20}$ &	$(2.17\pm0.36)\times10^{-6}$	&	\nodata	&	18	&	\\
050819	&	\nodata	&	$36\pm4$	&	$(4.2\pm0.8)\times10^{-7}$	&	\nodata	&	19	&	\\
050820A	&	2.612	&	$26\pm2$	&	$(1.9\pm0.2)\times10^{-6}$	&	20	&	21	&	\\
050822	&	\nodata	&	$102\pm2$	&	$(3.4\pm0.3)\times10^{-6}$	&	\nodata	&	22	&	\\
{\it 050826}\tablenotemark{c}	&\nodata&	$35\pm8$	&	$(4.3\pm0.7)\times10^{-7}$	&	\nodata	&	23	&	\\
050904	&	6.29	&	$225\pm10$	&	$(5.4\pm0.2)\times10^{-6}$	&	24	&	25	&	\\
050908	&	3.3437	&	$20\pm2$	&	$(5.1\pm0.5)\times10^{-7}$	&	26	&	27	&	\\
050915A	&	\nodata	&	$53\pm3$	&	$(8.8\pm0.9)\times10^{-7}$	&	\nodata	&	28	&	\\
050916	&	\nodata	&	$90\pm10$	&	$(1.1\pm0.4)\times10^{-6}$	&	\nodata	&	29	&	\\
050922B	&	\nodata	&	$80\pm10$	&	$(1.8\pm0.3)\times10^{-6}$	&	\nodata	&	30	&	\\
{\it 051016B\tablenotemark{d}}	&0.936	&	$4.0\pm0.1$	&	$(1.7\pm0.2)\times10^{-7}$	&	31	&	32	&	\\
051117A	&	\nodata	&	$140\pm10$	&	$(4.6\pm0.16)\times10^{-6}$	&	\nodata	&	33	&	\\
051210	&	\nodata	&	$1.4\pm0.2$	&	$(8.3\pm1.4)\times10^{-8}$	&	\nodata	&	34	& Short	\\
051227	&	\nodata	&	$8.0\pm0.2$	&	$(2.3\pm0.3)\times10^{-7}$	&	\nodata	&	35	& 	 \\

060108	&	\nodata	&	$14.4\pm1$	&	$(3.7\pm0.4)\times10^{-7}$	&	\nodata	&	36	&	\\
{\it 060109}\tablenotemark{e}	&\nodata&	$116\pm3$	&	$(6.4\pm1.0)\times10^{-7}$	&	\nodata	&	37	&	\\
060111A	&	\nodata	&	$13\pm1$	&	$(1.18\pm0.05)\times10^{-6}$	&	\nodata	&	38	&	\\
060115	&	3.53	&	$142\pm5$	&	$(1.8\pm0.2)\times10^{-6}$	&	39	&	40	&	\\
060124\tablenotemark{f}	&	2.296	&	$321\pm2$	&	$(1.40\pm0.03)\times10^{-5}$	&	41	&	42	& 	
   \enddata 
    \tablenotetext{a}{GRBs with number in italic were considered for their behaviour, but did not 
	        offer sufficiently high statistics to allow full analysis (see \S\ref{flares:sample}).} 
    \tablenotetext{b}{Drawn form refined BAT GCN Circulars in the 15--150\,keV band.}
    \tablenotetext{c}{A low-signal late-time flare is observed and no analysis was performed.}
    \tablenotetext{d}{A flattening in the XRT light curve is observed starting from $t \sim 200$s and lasting through the first SAA data gap. 
	A fit with a Gaussian centered at $t \sim 650$s provides a significantly worse fit than a combination of power laws, 
	hence this event was not included in the restricted sample.}
    \tablenotetext{e}{A flattening in the XRT light curve is observed starting from $t \sim 10^3$s and lasting through the first SAA data gap.}
    \tablenotetext{f}{As reported in \citet{Romano2006:060124}, a separate fit was performed to the prompt and the afterglow parts of the X-ray
			light curve. Here we do not consider the spikes in the prompt.}
    \tablerefs{
(1) \citet{Krimm2005:050406fluences}; 
(2) \citet{Sakamoto2005:050421fluences}; 
(3) \citet{Cummings2005:050502Bfluences}; 
(4) \citet{Retter2005:050607fluences}; 
(5) \citet{Markwardt2005:050712fluences}; 
(6) \citet{Palmer2005:050713Afluences}; 
(7) \citet{Tueller2005:050714Bfluences};
(8) \citet{Barthelmy2005:050716fluences};
(9) \citet{Prochaska2005:050724redshift};
(10) \citet{Krimm2005:050724fluences};
(11) \citet{Barthelmy2005:050726fluences};
(12) \citet{Chen2005:050730redshift};
(13) \citet{Markwardt2005:050730fluences};
(14) \citet{Sakamoto2005:050801fluences};
(15) \citet{Palmer2005:050802fluences};
(16) \citet{Bloom2005:050803redshift};
(17) \citet{Parsons2005:050803fluences};
(18) \citet{Tueller2005:050814fluences};
(19) \citet{Barthelmy2005:050819fluences};
(20) \citet{Prochaska2005:050820Aredshift};
(21) \citet{Cummings2005:050820Afluences};
(22) \citet{Hullinger2005:050822fluences};
(23) \citet{Markwardt2005:050826fluences};
(24) \citet{Haislip2006Nat:050904redshift};
(25) \citet{Sakamoto2005:050904fluences};
(26) \citet{Fugazza2005:050908redshift};
(27) \citet{Sato2005:050908fluences};
(28) \citet{Barthelmy2005:050915Afluences};
(29) \citet{Fenimore2005:050916fluences};
(30) \citet{Hullinger2005:050922Bfluences};
(31) \citet{Soderberg2005:051016Bredshift};
(32) \citet{Barbier2005:051016Bfluences};
(33) \citet{Palmer2005:051117Afluences};
(34) \citet{Sato2005:051210fluences};
(35) \citet{Hullinger2005:051227fluences};
(36) \citet{Sakamoto2006:060108fluences};
(37) \citet{Palmer2006:060109fluences};
(38) \citet{Sato2006:060111Afluences};
(39) \citet{Piranomonte2006:060115redshift};
(40) \citet{Barbier2006:060115fluences};
(41) \citet{Mirabal2006:060124redshift};
(42) \citet{Romano2006:060124}.	
} 
  \end{deluxetable}


  \begin{deluxetable}{llclcl}
   \tablewidth{0pc} 	      	
   \tablecaption{Fits to the XRT light curves: continuum parameters.\label{flares:tab:cont_pars}}
   \tablehead{	
\colhead{GRB} & \colhead{$\alpha_1$\tablenotemark{a}} & 	\colhead{$t_{\rm b1}$} & 
\colhead{$\alpha_2$\tablenotemark{a}} & \colhead{$t_{\rm b2}$} & \colhead{$\alpha_3$\tablenotemark{a}} \\
\colhead{} & 	\colhead{} & 	\colhead{(s)} & 	\colhead{} & 	\colhead{(s)} & 	\colhead{}  \\
\colhead{(1)} 	 & \colhead{(2)} 	 & \colhead{(3)}  & \colhead{(4)}  & \colhead{(5)}  & \colhead{(6)} } 	
\startdata
050406    & $1.58_{-0.17}^{+0.17}$ & $(4.36^{+6.23}_{-0.53})\times 10^3$ &  $0.50_{-0.14}^{+0.13}$ & \nodata 	     & \nodata          \\
050421	  & $3.10_{-0.09}^{+0.11}$ & \nodata                  		 &  \nodata                & \nodata 	     & \nodata          \\
050502B   & $0.75_{-0.04}^{+0.04}$ & $(15.2_{-4.3}^{+5.2})\times10^{4} $ &  $1.77^{+0.32}_{-0.26}$ & \nodata                  & \nodata       \\
050607	  & $1.65_{-0.16}^{+0.17}$ & $ 1.45\times 10^3\, \tablenotemark{b}$ &  $0.52^{+0.14}_{-0.16}$ & $1.54\times 10^4\, \tablenotemark{b} $& $ 1.34^{+0.39}_{-0.26} $  \\
050712	  & $2.17^{+0.38}_{-0.67}$ & $ 3.44\times 10^2\, \tablenotemark{b}$ &  $3.12^{+0.37}_{-0.25}$ & $8.39\times 10^2\, \tablenotemark{b} $& $ 0.43^{+0.27}_{-0.29} $  \\
050713A	  & $7.16^{+0.84}_{-0.68}$ & $ 1.12\times 10^2\, \tablenotemark{b}$ &  $0.81\, \tablenotemark{b}$& \nodata 	                          & \nodata                \\
050714B	  & $6.79^{+0.35}_{-0.38}$ & $(3.90_{-0.3}^{+0.31})\times10^2  $ &  $0.49^{+0.10}_{-0.09}$ & $(8.03_{-7.80}^{+7.81})\times 10^4 $ & $ 0.79^{+0.33}_{-0.32}$  \\
050716	  & $1.32^{+0.02}_{-0.07}$ & $(4.70_{-0.15}^{+0.05})\times10^2 $ &  $8.8^{+1.40}_{-1.60}$  & \nodata 	                          & \nodata                  \\ 
050724	  & $1.53^{+0.07}_{-0.07}$ & $(1.90_{-0.04}^{+0.05})\times10^2 $ &  $5.8^{+0.50}_{-0.30}$  & $(5.53_{-0.54}^{+0.60})\times 10^2 $ & $ 0.78_{-0.18}^{+0.13} $  \\ 
050726	  & $0.95^{+0.04}_{-0.03}$ & $(8.53_{-1.43}^{+1.32})\times10^3 $ &  $1.89^{+0.16}_{-0.20}$ & \nodata 	                          & \nodata                 \\ 
050730    & $0.28^{+0.04}_{-0.09}$ & $(5.52_{-0.26}^{+0.34})\times10^3 $ &  $1.97^{+0.06}_{-0.04}$ & \nodata 	                          & \nodata                 \\ 
050801	  & $0.57^{+0.22}_{-0.16}$ & $(4.67_{-1.87}^{+1.90})\times10^2 $ &  $1.24^{+0.09}_{-0.08}$ & \nodata 	                          & \nodata                  \\ 
050802	  & $0.27\, \tablenotemark{b}$ &$(8.80_{-1.51}^{+2.10})\times10^3 $ &  $1.60^{+0.19}_{-0.25}$ & \nodata 	                  & \nodata            \\ 
050803	  & $4.54^{+0.26}_{-0.29}$ & $(4.46_{-0.32}^{+0.33})\times10^2 $ &  $0.03^{+0.01}_{-0.08}$ & $(1.27_{-0.58}^{+0.58})\times 10^4  $& $ 1.59^{+0.03}_{-0.04} $  \\
050814	  & $3.26^{+0.12}_{-0.20}$ & $(9.99_{-0.98}^{+0.59})\times10^2 $ &  $0.56^{+0.09}_{-0.14}$ & $(8.46_{-1.26}^{+1.24})\times 10^4  $& $ 2.44^{+0.34}_{-0.46} $  \\
050819	  & $3.22\, \tablenotemark{b}$& $(8.18\times10^2\, \tablenotemark{b} $ &  $0.27\, \tablenotemark{b}$ & \nodata	                & \nodata                 \\
050820A	  & $2.25^{+0.14}_{-0.17}$ & $(2.00_{-0.19}^{+0.14})\times10^2 $ &  $0.03\, \tablenotemark{b}$& $(4.79_{-0.34}^{+0.52})\times 10^3  $& $1.27^{+0.05}_{-0.06} $  \\
050822	  & $2.99\, \tablenotemark{b}$& $7.50\times10^2\, \tablenotemark{b}  $ &  $0.40\, \tablenotemark{b}$& $(2.22\times 10^4 \, \tablenotemark{b} $& $1.72\, \tablenotemark{b}$ \\
050904	  & $1.57^{+0.12}_{-0.13}$ & $(3.35_{-0.42}^{+0.40})\times10^2 $ &  $2.26^{+0.11}_{-0.10}$ & $(1.70_{-0.36}^{+0.49})\times 10^4 $& $0.50\, \tablenotemark{b} $\\
050908	  & $1.12^{+0.06}_{-0.06}$ & \nodata                  		 &  \nodata                & \nodata	                & \nodata                    \\
050915A	  & $0.42^{+0.28}_{-0.27}$ & $(1.74_{-0.71}^{+2.36})\times10^3 $ &  $1.20^{+0.20}_{-0.10}$ & \nodata	                & \nodata                    \\
050916	  & $0.95^{+0.30}_{-0.25}$ & \nodata	             		 &  \nodata                & \nodata	                & \nodata                    \\
050922B	  & $3.33^{+0.37}_{-0.30}$ & \nodata	             		 &  \nodata                & \nodata	                & \nodata                    \\
051117A	  & $0.66^{+0.11}_{-0.10}$ & \nodata                  		 &  \nodata                & \nodata 	                & \nodata                    \\
051210    & $2.58^{+0.25}_{-0.17}$ & \nodata	             		 &  \nodata                & \nodata	                & \nodata                    \\
051227 	  & $2.50^{+0.15}_{-0.15}$ & $7.37\times10^2\, \tablenotemark{b}$   &  $0.18\, \tablenotemark{b}$& $3.10\times 10^3\, \tablenotemark{b}    $& $1.22\, \tablenotemark{b} $  \\ 
060108	   & $2.60^{+0.55}_{-0.55}$ & $(2.54_{-0.46}^{+0.45})\times10^2 $ &  $0.37^{+0.05}_{-0.05}$ & $(1.87_{-0.36}^{+0.36}) \times 10^3 $& $1.22^{+0.09}_{-0.09} $   \\
060111A	   & $-4.25_{-0.44}^{+0.36}$& $(3.25_{-0.37}^{+0.28})\times10^2 $ &  $6.26^{-0.27}_{+0.28}$ & $(7.38_{-0.33}^{+0.37}) \times 10^2 $& $0.90_{-0.05}^{+0.05} $  \\
060115     & $3.29^{+0.21}_{-0.29}$ & $(5.74_{-0.56}^{+0.86})\times10^2 $ &  $0.70^{+0.05}_{-0.08}$ & $(3.91_{-1.37}^{+2.13}) \times 10^4 $& $1.31^{+0.22}_{-0.20} $     \\ 
060124\tablenotemark{c}	& $ 0.44_{-0.08}^{+0.07}$ 			 & $(1.0$--$11.5)\times 10^3$ & $1.21\pm0.04$ & $(1.05_{-0.14}^{+0.17})\times 10^{5}$ & $1.58\pm0.06$   \\
\enddata
\tablenotetext{a}{These slopes do not strictly correspond to phases I, II, and III of the canonical XRT light curve.}  
\tablenotetext{b}{Parameter fixed.}  
\tablenotetext{c}{The fits of prompt (first orbit) and afterglow were performed separately. 
The first break ($t_{\rm b1}$) is not well defined, since it occurs during a SAA passage, that lasts from $\sim 1000$ to $\sim 11500$\,s.} 
\end{deluxetable} 
\clearpage 


  \begin{deluxetable}{lccccrrrrr}
   \tablewidth{0pc} 	      	
  \tabletypesize{\scriptsize}
   \tablecaption{Fits to the XRT flares: Gaussians, power-laws and exponentials.\label{flares:tab:fits}}
   \tablehead{	
\colhead{GRB} & \colhead{} & \colhead{Gaussian}  & \colhead{} &  \colhead{$\Delta F/F$}  & \colhead{EW}  
	& \colhead{$\alpha_{\rm fall}$}  & \colhead{$\tau_{90}$}  & \colhead{$\Delta t_{\rm fall}/\Delta t_{\rm rise}$} 
	& \colhead{$\Delta t/t$\tablenotemark{a}}   \\
\colhead{} & \colhead{Center (s)} & \colhead{Width (s)}    & \colhead{Norm (count s$^{-1}$)}  & \colhead{} & \colhead{(s)}  & \colhead{}
					& \colhead{(s)}    & \colhead{}   & \colhead{}   \\
\colhead{(1)}  & \colhead{(2)}     & \colhead{(3)}  & \colhead{(4)}  & \colhead{(5)}  & \colhead{(6)}
	   & \colhead{(7)} & \colhead{(8)} & \colhead{(9)}  & \colhead{(10)} 
}
\startdata
 050406	 &$211_{-5}^{+5}	      $&$17.9_{-4.6}^{+12.3}	    $&$	4.6_{-1.3}^{+1.2} $               &$7.71$  & 686	&\nodata& 184.0  &1.520  &0.882  \\ 
 050421	 &$111_{-2}^{+0}	      $&$1.7_{-0.1}^{+0.1}	    $&$	191.2_{-107.8}^{+1186.8}$         &$12.35$ & \nodata &\nodata& \nodata & \nodata & \nodata  \\ 
          &$154_{-3}^{+3}	      $&$6.2_{-4.1}^{+4.3}	    $&$	4.7_{-1.9}^{+5.4}	$         &$0.84$  & \nodata &\nodata& \nodata & \nodata & \nodata  \\ 
050502B	 &$719_{-2}^{+1}	      $&$100.1_{-1.02}^{+1.2}	    $&$	88.0_{-1.4}^{+1.3}	$         &$38.55$ & 127320	&$ 6.3\pm0.38$  & 523.6  &1.450  &1.352   \\
          &$33431_{-3391}^{+4057}      $&$6273.0_{-2258}^{+2815}     $&$	0.012_{-0.005}^{+0.006}	$ &$0.28$  & \nodata	&\nodata& \nodata & \nodata & \nodata  \\ 
          &$74637_{-2581}^{+2429}      $&$26742.9_{-2440.9}^{+2761.1}$&$	0.027_{-0.003}^{+0.003}	$ &$1.57$  & 432630	&$ 4.67\pm0.34$ & \nodata & \nodata & \nodata   \\
050607	 &$330_{-7}^{+8}	      $&$36.1_{-5.4}^{+5.5}	    $&$	15.8_{-3.1}^{+3.3}	$         &$21.84$ & 1813	&$ 3.39\pm0.24$ & 266.5  &1.610  &0.798  \\ 
050712	 &$245.7_{-5.7}^{+3.3}        $&$31.1_{-6.6}^{+5.0}         $&$7.3_{-1.2}^{+1.2}$	          &$1.37$  &\nodata	&\nodata\tablenotemark{a}& \nodata & \nodata  & \nodata  \\
          &$486.1_{-3.7}^{+4.9}        $&$16.7_{-2.8}^{+4.1}         $&$7.4_{-1.6}^{+1.5}$	          &$5.10$  & 165	&$ 2.87\pm0.41$ & \nodata & \nodata & \nodata   \\
          &$913.5_{-21.1}^{+21.8}      $&$98.9_{-15.7}^{+29.3}       $&$1.0_{-0.27}^{+0.47}$	          &$6.35$  & \nodata	&\nodata& \nodata & \nodata & \nodata  \\ 
050713A	 &$112.2_{-0.5}^{+0.6}        $&$5.9_{-0.5}^{+0.5}          $&$170.5_{-18.3}^{+17.6}$	          &$19.77$ & 190	&$ 2.92\pm0.25$ & 49.5   &3.070  &0.445  \\ 
          &$173.4_{-1.5}^{+1.6}        $&$16.3_{-2.1}^{+2.0}         $&$23.5_{-2.2}^{+2.4}$	          &$3.97$  & 94 	&$ 3.1\pm0.2  $ & 82.9   &2.790  &0.494  \\ 
          &$399.8_{-5.3}^{+9.9}        $&$23.3_{-3.6}^{+5.1}         $&$24.9_{-2.16}^{+3.4}$	          &$8.27$  & \nodata	&\nodata & \nodata & \nodata & \nodata  \\ 
          &$126.2_{-3.1}^{+3.8}        $&$10.8_{-1.7}^{+1.3}         $&$55.6_{-9.6}^{+11.5}$	          &$7.26$  & \nodata	&\nodata& \nodata & \nodata & \nodata  \\ 
050714B	 &$399_{-8}^{+8}              $&$52.4_{-6.1}^{+9.4}	    $&$	4.4_{-0.9}^{+0.8}	$         &$69.86$ & 2313	&$ 3.2\pm0.43 $& 344.8   &3.410  &0.928  \\ 
050716	 &$175_{-66}^{+0}	      $&$48_{-15}^{+19}	            $&$	9.0_{-2}^{+14}		$         &$0.48$  & 240	&$ 0.51\pm0.24$& 622.0   &4.750  &3.514  \\ 
          &$382_{-6}^{+5}	      $&$16.3_{-5.6}^{+7.2}	    $&$	3.8_{-1.3}^{+1}	$ 	          &$0.57$  &  383	&$ 2.13\pm0.51$& 482.9   &2.700  &1.283  \\ 
050724	 &$275_{-5}^{+5}              $&$30.6_{-6}^{+6.6}	    $&$	7.2_{-1.1}^{+1.1}	$         &$1.35$  & 84 	&$ 2.52\pm0.5 $& \nodata & \nodata & \nodata \\
          &$327_{-9}^{+6}              $&$12.7_{-5.0}^{+6.3}	    $&$	3.1_{-1.0}^{+1.0}	$         &$1.58$  & 67 	&$ 4.43\pm0.8 $& \nodata & \nodata & \nodata \\
          &$(5.7_{-0.3}^{+0.2})\times10^4$&$(1.9_{-0.3}^{+0.3})\times10^{4}$& $	0.030_{-0.003}^{+0.004}	$ &$11.99$ & 737109	&$ 3.13\pm0.19$& 112365 &1.720   &2.045 \\
050726	 &$168_{-5}^{+5}	      $&$8.2_{-4.4}^{+6.7}	    $&$	3.1_{-1.6}^{+1.9}	$         &$0.46$  & 8  	&$ 3.7\pm1.2  $& 33.0	 &0.492  &0.199 \\
          &$273_{-4}^{+4}	      $&$27.0_{-4.4}^{+5.1}	    $&$	6.6_{-1}^{+1}		$         &$1.57$  & 126	&$ 3.5\pm0.53 $& 122.0   &1.120  &0.446 \\
050730	 &$131.8_{-59.6}^{+12.7}      $&$32.7_{-8.3}^{+24.4}        $&$8.3_{-1.4}^{+9.2}$	          &$1.33$  & \nodata	&\nodata& \nodata & \nodata & \nodata \\
          &$234.2_{-2.4}^{+2.7}        $&$14.5_{-2.8}^{+3}           $&$5.3_{-0.8}^{+0.9}$		  &$1.00$  & 43 	&$ 4.9\pm1.1  $& \nodata & \nodata & \nodata \\
          &$436.5_{-2.2}^{+1.5}        $&$38.5_{-2.5}^{+2.8}         $&$9.0_{-0.5}^{+0.6}$		  &$2.02$  & 370	&\nodata& \nodata & \nodata & \nodata \\
          &$685.8_{-2.7}^{+2.8}        $&$23.8_{-3.5}^{+3.9}         $&$5.19_{-0.61}^{+0.69}$	          &$1.32$  &224 	&\nodata& \nodata & \nodata & \nodata \\
          &$742  		      $&$ 10 		            $&$3.0    $				  &$0.78$  & \nodata	&\nodata& \nodata & \nodata & \nodata \\
          &$4526.2_{-107.2}^{+112.8}   $&$408.1_{-94.1}^{+126.9}     $&$0.86_{-0.26}^{+0.24}$ 	          &$0.37$  &350 	&\nodata& \nodata & \nodata & \nodata \\
          &$10223.6_{-477.6}^{+203.4}  $&$847_{-179}^{+3}            $&$0.87_{-0.10}^{+0.12}$ 	          &$1.34$  &1897	&\nodata& \nodata & \nodata & \nodata \\
          &$12182.9       	      $&$383.2     		    $&$0.4 $			          &$0.87$  & \nodata	&\nodata& \nodata & \nodata & \nodata \\
050801   &$284_{-35}^{+48}	      $&$49.5_{-42.5}^{+42.5}	    $&$1.0_{-0.7}^{+0.9}	$	  &$0.91$  & \nodata	&\nodata& \nodata & \nodata & \nodata \\
050802	 &$464_{-31}^{+31}	      $&$100_{-40}^{+33}	    $&$2.14_{-0.74}^{+0.46}$ 		  &$2.25$  &159 	&$ 2.54\pm0.35$& 926.3   &5.300  &2.327 \\
050803	 &$332_{-19}^{+19}            $&$29.0_{-22}^{+22}           $&$0.8_{-0.5}^{+0.5}	$         &$0.85$  &65  	&$ 1.7\pm0.8  $& \nodata & \nodata &\nodata \\
          &$604                        $&$189.2                      $&$1.00$		            	  &$4.05$  &357 	&$ 3.1\pm1.2  $& \nodata & \nodata & \nodata \\
          &$1201		              $&$164.2                      $&$0.67$ 			  &$2.66$  &404 	&$ 1.3\pm3.2  $& \nodata & \nodata & \nodata \\
050814	 &$2286_{-127}^{+769}         $&$299.0_{-118.9}^{+423.7}    $&$0.12_{-0.02}^{+0.02}$	          &$1.30$  & \nodata	&\nodata& \nodata & \nodata & \nodata \\
050819	 &$177_{-19}^{+7}	      $&$13.9_{-5.5}^{+11.72}	    $&$2.1_{-0.8}^{+1.0}	$ 	  &$0.67$  &8		&\nodata& \nodata & \nodata & \nodata \\
050820A	 &$241_{-1}^{+0}	      $&$9.5_{-0.2}^{+0.3}	    $&$	231.0_{-6.2}^{+6.2}	$	  &$77.45$ & \nodata	&\nodata & \nodata & \nodata & \nodata \\
050822	 &$142.7_{-1.1}^{+1.2}        $&$15.2_{-0.8}^{+1.0}           $&$54.7_{-3.5}^{+3.7}$		  &$1.09$  &59  	&$ 4.34\pm0.17$& \nodata & \nodata & \nodata \\
          &$241.8_{-1.6}^{+1.9}        $&$12.4_{-1.7}^{+1.7}         $&$15.5_{-2.1}^{+2.3}$		  &$1.50$  &129 	&$ 2.78\pm0.17$& 110.3   &1.280  &0.459 \\
          &$465.7_{-1.6}^{+1.6}        $&$49.0_{-0.4}^{+2.3}         $&$43.5_{-1.46}^{+1.38}$		  &$29.89$ &6851	&$ 5.06\pm0.18$& 328.0   &0.630  &0.708 \\
050904	 &$448.6_{-4}^{+3.7}          $&$45.9_{-3.8}^{+4.5}         $&$20.7_{-1.4}^{+1.2}$	          &$2.22$  &401 	&$ 4.52\pm0.32$& \nodata & \nodata & \nodata \\
          &$975.5_{-32.5}^{+38.5}      $&$62.8_{-32.5}^{+36.9}       $&$ 1.0_{-0.2}^{+0.5}$	          &$0.62$  &162 	& \nodata  & \nodata & \nodata & \nodata \\	 
          &$1265.5_{-27.0}^{+28.0}     $&$81.6_{-28.2}^{+30.1}       $&$ 1.1_{-0.3}^{+0.4}$	          &$1.23$  &364 	& \nodata  & \nodata & \nodata & \nodata \\	 
          &$7112.8_{-102.8}^{+147.2}   $&$790.4_{-81.6}^{+103.1}     $&$1.6_{-0.1}^{+0.1}$	          &$88.46$ & \nodata	& \nodata& \nodata & \nodata & \nodata \\
          &$16682.2_{-263.2}^{+260.8}  $&$3194.8_{-227.8}^{+212.2}   $&$0.77_{-0.04}^{+0.05}$	          &$292.29$& \nodata	& \nodata& \nodata & \nodata & \nodata \\
          &$31481.1_{-762.1}^{+724.9}  $&$7149.6_{-647.6}^{+687.4}   $&$0.31_{-0.02}^{+0.02}$	          &$166.67$& \nodata	& \nodata& \nodata & \nodata & \nodata \\
050908	 &$146_{-18}^{+10}	      $&$23_{-23}^{+-23}	    $&$2.17_{-0.97}^{+0.93}$	          &$1.72$  & 88      &\nodata\tablenotemark{b}& \nodata & \nodata & \nodata  \\
          &$425_{-11}^{+18}	      $&$45_{-15}^{+18}	  $&$2.4_{-0.7}^{+1.1}	$ 			  &$6.29$  & 1132	&$ 2.36\pm0.11$& 295.6   &2.660  &0.727 \\
050915A	 &$107_{-5}^{+2}	      $&$15.5_{-2.6}^{+5.6}	    $&$	12.20_{-1.5}^{+1.5}	$	  &$13.36$ &43  	&$ 3.35\pm0.38$& \nodata & \nodata & \nodata  \\
050916	 &$18750_{-105}^{+236}        $&$425_{-105}^{+205} 	    $&$	0.2_{-0.1}^{+0.1}	$         &$25.22$ &130717	&\nodata & \nodata & \nodata & \nodata \\
          &$21463_{-425}^{+696}	      $&$2222_{-360}^{+592}	    $&$0.1_{-0.03}^{+0.04}	$         &$14.34$ & \nodata	&\nodata& \nodata & \nodata & \nodata \\
050922B	 &$375_{-1}^{+2}	      $&$9.2_{-1.7}^{+2.1}	    $&$	23.0_{-4}^{+3}		$	  &$0.97$  &221 	&$ 1.66\pm0.33$& 175.4   &1.330  &0.466 \\
          &$490_{-8}^{+8}	      $&$37.7_{-8}^{+9 }	    $&$	 6.7_{-1.3}^{+1.1}	$         &$0.69$  &410 	& \nodata      & 254.7   &2.020  &0.508 \\
          &$858_{-9}^{+10}	      $&$123_{-8}^{+9}		    $&$22.0_{-2}^{+2}	$	          &$14.64$ &14336	&$ 6.76\pm0.42$& 464.9   &1.420  &0.572 \\
051117A	 &$132_{-5}^{+5}              $&$48_{-4}^{+4}               $&$102_{-9}^{+6}$			  &$3.23$  & 27      &$2.72\pm0.46$&$ 331.9$& 2.5 & 2.192 \\
          &$376_{-14}^{+18}            $&$203_{-20}^{+14}             $&$47_{-6}^{+4}$			  &$3.01$  & 195     &\nodata& \nodata & \nodata & \nodata \\
          &$955_{-6}^{+7}              $&$69_{-5}^{+6}               $&$29_{-2}^{+1}$			  &$3.41$  & 395     &$3.51\pm 0.28$& \nodata & \nodata & \nodata \\
          &$1110_{-5}^{+5}             $&$50_{-4}^{+4}               $&$27_{-2}^{+2}$			  &$3.51$  & 1201    &\nodata& \nodata & \nodata & \nodata \\
          &$1341_{-2}^{+3}             $&$43_{-2}^{+3}               $&$49_{-2}^{+3}$		   	  &$7.21$  & \nodata &\nodata& 603.4	 &8.060  &0.453 \\
          &$1516_{-7}^{+9}             $&$135_{-9}^{+7}              $&$30_{-1.3}^{+1.2}$		  &$4.78$  & \nodata &$6.6\pm4.0$& \nodata & \nodata & \nodata \\
051210	 &$134.4_{-4.4}^{+3.6}        $&$10.4_{-4.1}^{+5.7}	    $&$4.7_{-1.7}^{+1.7}	$	  &$1.18$  &16       &$ 4.05\pm0.57$& 49.2	 &0.490  &0.360 \\
          &$216.2                      $&$63.1    	            $&$0.62	$ 		          &$0.53$  & \nodata &\nodata & \nodata & \nodata & \nodata \\
051227	 &$124.2                      $&$10.5                       $&$5.15	$		          &$0.88$  & \nodata &$ 2.05\pm0.5 $& \nodata & \nodata & \nodata \\
060108	 &$303.5_{-24.5}^{+23.5}      $&$44.5_{-30.5}^{+125.5}	    $&$0.3_{-0.12}^{+0.18}$		  &$1.83$  &25       &$ 	      $& \nodata & \nodata & \nodata \\
060111A	 &$95.1_{-1.5}^{+1.4}         $&$22.8_{-1.7}^{+2.1}         $&$67.6_{-2.8}^{+2.7}$	          &$165.51$&73       &$ 3.53\pm0.39$& 144.4	 &0.800  &1.405 \\
  	 &$166.9_{-1.6}^{+1.6}        $&$18.4_{-1.8}^{+2.0}           $&$34.7_{-2.2}^{+2.1}$	          &$7.78$  &54       &$ 4.5\pm1.2  $&	 120.7   &1.230  &0.719 \\
   	 &$280.1_{-1.5}^{+1.3}        $&$20.6_{-1.79}^{+1.73}       $&$85.0_{-5.24}^{+4.59}$		  &$2.11$  &931      &$ 6.51\pm0.4 $&	 177.5   &2.430  &0.620 \\
060115   &$431.9_{-18.5}^{+18.5}      $&$79.1_{-23.5}^{+25.8}       $&$1.91_{-0.45}^{+0.54}$         	  &$2.53$  &144.2    & \nodata & \nodata & \nodata  & \nodata
   \enddata
	\tablenotetext{a}{Using $\Delta t=\tau_{90}$ (the time defined in terms of $f=0.05$, \S\ref{flares:alltaus}) and $t=t_{\rm peak}$.} 
	\tablenotetext{b}{GRB~050712 and GRB~050908 have a first flare that quite likely is part of the prompt emission.  
         In addition the decay does not show a very high statistics. 
	} 
    \end{deluxetable}
    \clearpage

 \begin{deluxetable}{lcccrr}
  \tablewidth{0pc} 	      	
  \tablecaption{Properties of the 46 gamma-ray pulses detected from the BAT light curves of
28 GRBs with X-ray flares.\label{flares:tab:bat_peaks}} 
  \tablehead{	
\colhead{GRB}  & \colhead{N}   	&\colhead{Bin} 	& \colhead{Peak}&\colhead{Peak} & \colhead{Error on Peak} \\
\colhead{Name} & \colhead{}  	&\colhead{T (s)}& \colhead{Time (s)}&\colhead{Rate (counts s$^{-1}$)} & \colhead{Rate (counts s$^{-1}$)} \\
\colhead{(1)}  & \colhead{(2)} 	&\colhead{(3)} & \colhead{(4)} 	&\colhead{(5)} 	& \colhead{(6)}  
}
  \startdata	
050406	&	1&	2.176	&	2.24	&	0.04144	&	0.00546	\\
050421	&	1&	10.688	&	11.072	&	0.01538	&	0.00256	\\
050502B	&	1&	0.384	&	0.864	&	0.23168	&	0.01581	\\
050607	&	1&	1.216	&	1.656	&	0.12026	&	0.00919	\\
 	&	2&	13.952	&	16.888	&	0.03721	&	0.00256	\\
050712	&	1&	30.464	&	26.912	&	0.03095	&	0.00237	\\
050713A	&	1&	5.312	&	$-$54.89&	0.04249	&	0.00675	\\
 	&	2&	1.664	&	2.712	&	0.53963	&	0.01724	\\
 	&	3&	6.400	&	10.84	&	0.40561	&	0.00926	\\
 	&	4&	3.520	&	69.4	&	0.05585	&	0.00542	\\
 	&	5&	12.032	&	116.70	&	0.01849	&	0.00233	\\
050714B	&	1&	23.68	&	52.392	&	0.02617	&	0.00313	\\
050716	&	1&	4.288	&	11.81	&	0.22210	&	0.01168	\\
 	&	2&	11.648	&	46.94	&	0.12358	&	0.00648	\\
050724	&	1&	0.128	&	0.104	&	1.29947	&	0.07964	\\
 	&	2&	0.064	&	210.92	&	0.29937	&	0.04221	\\
050726	&	1&	2.368	&	$-$173.87&	0.09576	&	0.01595	\\
 	&	2&	7.104	&	7.89	&	0.10490	&	0.00968	\\
050730	&	1&	13.824	&	17.432	&	0.04678	&	0.00286	\\
050801	&	1&	0.512	&	0.592	&	0.20882	&	0.01472	\\
050802	&	1&	12.736	&	13.90	&	0.02504	&	0.00256	\\
050803	&	1&	1.600	&	1.168	&	0.32463	&	0.02357	\\
050814	&	1&	16.512	&	18.856	&	0.04661	&	0.00450	\\
050819	&	1&	16.128	&	23.224	&	0.02246	&	0.00231	\\
050822	&	1&	3.264	&	3.912	&	0.15719	&	0.01272	\\
 	&	2&	0.96	&	48.52	&	0.24216	&	0.01586	\\
 	&	3&	4.352	&	60.04	&	0.10462	&	0.00577	\\
 	&	4&	2.624	&	103.56	&	0.04600	&	0.00684	\\
050904	&	1&	6.976	&	29.768	&	0.06191	&	0.00493	\\
 	&	2&	15.808	&	125.128	&	0.05769	&	0.00229	\\
050908	&	1&	3.072	&	3.776	&	0.07615	&	0.00586	\\
050915A	&	1&	5.376	&	5.496	&	0.05597	&	0.00436	\\
 	&	2&	0.768	&	14.584	&	0.11584	&	0.01225	\\
 	&	3&	1.920	&	44.6	&	0.03813	&	0.00520	\\
050916	&	1&	16.448	&	51.304	&	0.03522	&	0.00376	\\
050922B	&	1&	15.168	&	52.072	&	0.06915	&	0.00699	\\
 	&	2&	14.336	&	103.4	&	0.05754	&	0.00705	\\
 	&	3&	1.408	&	263.464	&	0.06536	&	0.00757	\\
 	&	4&	1.216	&	271.656	&	0.06913	&	0.00825	\\
051117A	&	1&	8.704	&	11.264	&	0.08273	&	0.00388	\\
051210	&	1&	0.640	&	0.88	&	0.1057	&	0.01215	\\
051227	&	1&	0.640	&	0.80	&	0.12975	&	0.01213	\\
060108	&	1&	2.56	&	3.304	&	0.08358	&	0.00554	\\
060111A &       1&      2.816   &       5.792   &       0.19368 &       0.00639 \\
060115 	&	1&	5.312	&	6.352	&	0.05227	&	0.00438	\\
  	&	2&	3.52	&	98.192	&	0.09428	&	0.00513	
    \enddata 
   \end{deluxetable}  
   \clearpage

 \begin{deluxetable}{ccc}
  \tablewidth{0pc} 	      	
  \tablecaption{Frequency distribution of the number of BAT pulses vs.
the number of X-ray flares in 28 bursts.\label{flares:bat_xrt_pulses}} 
  \tablehead{	
\colhead{Number of BAT Pulses} & 	\colhead{Numer of X-ray flares} &    \colhead{Frequency} \\
\colhead{} & 	\colhead{} & 	    \colhead{} \\
\colhead{(1)} 	 & \colhead{(2)} 	 & \colhead{(3)} 
}
  \startdata	
1	&	1	&	8	\\
1	&	2	&	4	\\
1	&	3	&	4	\\
1	&	6	&	1	\\
1	&	8	&	1	\\
2	&	1	&	2	\\
2	&	2	&	2	\\
2	&	3	&	1	\\
2	&	6	&	1	\\
3       &       1       &       1       \\
4	&	3	&	2	\\
5	&	4	&	1	
    \enddata 
   \end{deluxetable}  
   \clearpage

 \begin{deluxetable}{lccccccccccc}
  \tablewidth{0pc} 
  \tabletypesize{\scriptsize}
  \tablecaption{Central times of the $\gamma$-ray pulses and X-ray flares for each GRB.
\label{flares:tab:xrtbat_times}} 
  \tablehead{	
\colhead{GRB} & \colhead{BAT} & \colhead{XRT} & \colhead{} & \colhead{} & \colhead{} & \colhead{} & \colhead{Times} & \colhead{} & \colhead{} & \colhead{} & \colhead{} \\
\colhead{Name} & \colhead{$n_{\gamma}$} & \colhead{$n_{\rm x}$} & \colhead{(s)} & \colhead{(s)} & \colhead{(s)} & \colhead{(s)} & 	\colhead{(s)} & \colhead{(s)} & \colhead{(s)} & \colhead{(s)} &   \colhead{(s)} \\
\colhead{(1)}  & \colhead{(2)} & \colhead{(3)} & \colhead{(4)} & \colhead{(5)} & \colhead{(6)} & \colhead{(7)} & \colhead{(8)} & \colhead{(9)} & \colhead{(10)} & \colhead{(11)} & \colhead{(12)} 
}
  \startdata	
050406	&	1&	1&	2.24	&	211.0	&	\nodata	&	\nodata	&	\nodata	&	\nodata	&	\nodata	&	\nodata	&	\nodata	\\
050421	&	1&	2&	11.072	&	111.0	&	154.0	&	\nodata	&	\nodata	&	\nodata	&	\nodata	&	\nodata	&	\nodata	\\
050502B	&	1&	3&	0.864	&	719.0	&	33431.	&	74637.	&	\nodata	&	\nodata	&	\nodata	&	\nodata	&	\nodata	\\
050607	&	2&	1&	1.656	&	16.888	&	330.0	&	\nodata	&	\nodata	&	\nodata	&	\nodata	&	\nodata	&	\nodata	\\
050712	&	1&	3&	26.912	&	245.7	&	486.1	&	913.5	&	\nodata	&	\nodata	&	\nodata	&	\nodata	&	\nodata	\\
050713A	&	5&	4&	$-$54.89&	2.712	&	10.84	&	69.4	&	116.70\tablenotemark{(a)}	&	112.2\tablenotemark{(a)}	&	126.2\tablenotemark{(a)}	&	173.4	&	399.8	\\
050714B	&	1&	1&	52.392	&	399.0	&	\nodata &       \nodata	&	\nodata	&	\nodata	&	\nodata	&	\nodata	&	\nodata	\\
050716	&	2&	2&	11.81	&	46.94	&	175.0	&	382.0	&	\nodata	&	\nodata	&	\nodata	&	\nodata	&	\nodata	\\
050724	&	2&	3&	0.104	&	210.92	&	275.0	&	327.0   &       57000.0	&	\nodata	&	\nodata	&	\nodata	&	\nodata	\\
050726	&	2&	2&	$-$173.87&	7.89	&	168.0	&	273.0	&	\nodata	&	\nodata	&	\nodata	&	\nodata	&		\\
050730	&	1&	8&	17.432	&	131.	&	234.2	&	436.5	&	685.8	&	742.0	&	4526.2	&	10223.6	       & 12182.9 \\
050801	&	1&	1&	0.592	&	284.0	&	\nodata	&	\nodata	&	\nodata	&	\nodata	&	\nodata	&	\nodata	&	\nodata	\\
050802	&	1&	1&	13.90	&	464.0	&	\nodata	&	\nodata	&	\nodata	&	\nodata	&	\nodata	&	\nodata	&	\nodata	\\
050803	&	1&	3&	1.168	&	332.0   &        604.0	&	1201.0	&	\nodata	&	\nodata	&	\nodata	&	\nodata	&	\nodata	\\
050814	&	1&	1&	18.856	&	2286.0	&	\nodata	&	\nodata	&	\nodata	&	\nodata	&	\nodata	&	\nodata	&	\nodata	\\
050819	&	1&	1&	23.224	&	177.0	&	\nodata	&	\nodata	&	\nodata	&	\nodata	&	\nodata	&	\nodata	&	\nodata	\\
050822	&	4&	3&	3.912	&	48.52	&	60.04	&	103.56	&	142.7	&	 241.8	&	465.7	&	\nodata	&	\nodata	\\
050904	&	2&	6&	29.768	&	125.128	&	448.6	&	975.5   &       1265.5  &        7112.8	&	16682.2	&	31481.1	&	\nodata	\\
050908	&	1&	2&	3.776	&	146.0	&	425.0	&	\nodata	&	\nodata	&	\nodata	&	\nodata	&	\nodata	&	\nodata	\\
050915A	&	3&	1&	5.496	&	14.584	&	44.6	&	107.0	&	\nodata	&	\nodata	&	\nodata	&	\nodata	&	\nodata	\\
050916  &       1&      2&      51.304  &       18750.0 &       21463.0 &       \nodata &       \nodata &       \nodata &       \nodata &       \nodata &       \nodata \\
050922B	&	4&	3&	52.072	&	103.4	&	263.464	&	271.656	&	375.0	&	490.0   &       858.0	&	\nodata	&	\nodata	\\
051117A	&	1&	6&	11.264	&	131.7   &        375.9  &         955.0 &       1110.0  &       1341.0  &       1516.0	&	\nodata	&	\nodata	\\
051210	&	1&	2&	0.88	&	134.4	&	216.2	&	\nodata	&	\nodata	&	\nodata	&	\nodata	&	\nodata	&	\nodata	\\
051227	&	1&	1&	0.80	&	124.2	&	\nodata	&	\nodata	&	\nodata	&	\nodata	&	\nodata	&	\nodata	&	\nodata	\\
060108	&	1&	1&	3.304	&	303.5	&	\nodata	&	\nodata	&	\nodata	&	\nodata	&	\nodata	&	\nodata	&	\nodata	\\
060111A &       1&      3&      5.792   &       95.1    &       166.9   &       280.1   &       \nodata &       \nodata &       \nodata &       \nodata &       \nodata \\
060115 	&	2&	1&	6.352	&	98.192	&	431.9	&	\nodata	&	\nodata	&	\nodata	&	\nodata	&	\nodata	&	\nodata	
    \enddata 
    \tablenotetext{(a)}{The BAT pulse occurs simultaneously with the two X-ray flares.} 
    \tablenotetext{(b)}{Errors on columns 4--12 are the binning times (see, \S\ref{flares:batxrt}).}
   \end{deluxetable}  
   \clearpage


\setcounter{figure}{0}  
\begin{figure*}	
	\epsscale{2.0}
	\vspace{-2truecm}
	\plotone{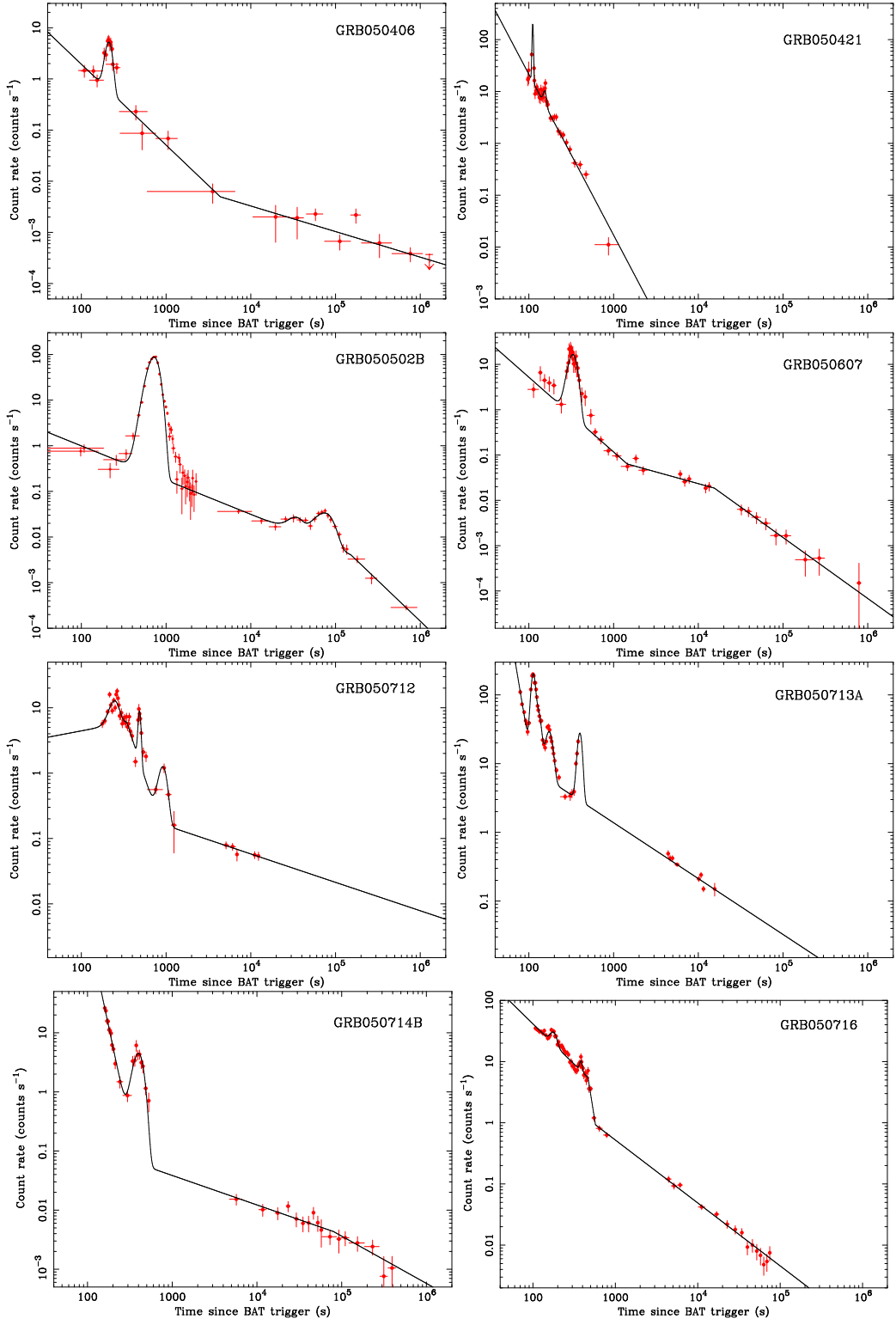}
	\caption[]{Flare fits. The thick line is the best fit to the XRT data 
	(filled circles) with a (multiply-)broken power law plus a number of 
	Gaussians (see Table~\ref{flares:tab:fits} for the fit parameters). 
	The continumm and Gaussian parameters are reported in Table~\ref{flares:tab:cont_pars} 
	and Table~\ref{flares:tab:fits}, respectively. For GRB~060124 we considered the prompt and
	afterglow portion of the light curve separately. 
	\label{flares:bigfig_1a}}
\end{figure*}
\clearpage 
\setcounter{figure}{0}
\begin{figure*}	
	\vspace{-2truecm}
	\plotone{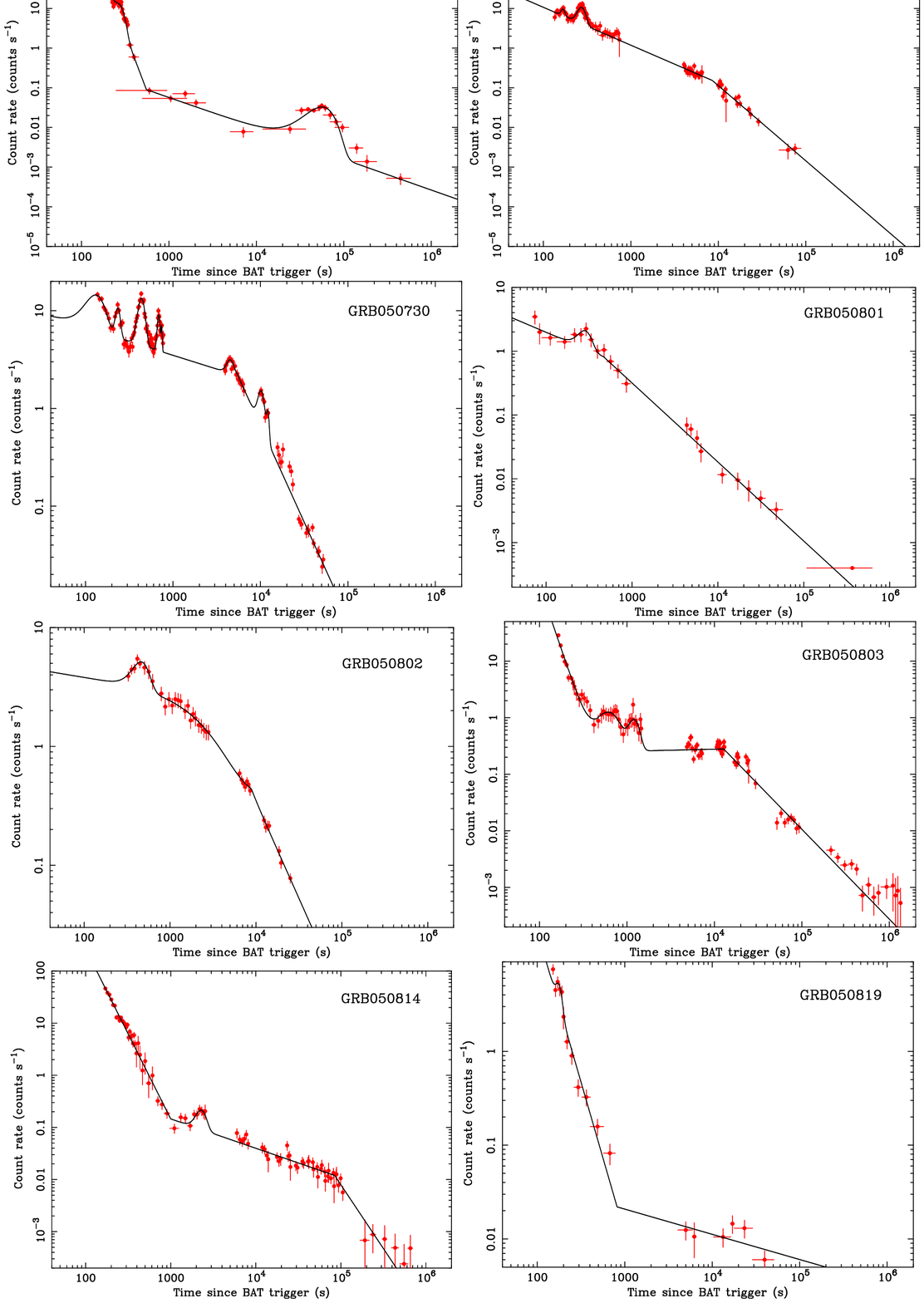}
	\caption[]{Continued.
	\label{flares:bigfig_1b}}
\end{figure*}
\clearpage 
\setcounter{figure}{0}
\begin{figure*}	
	\vspace{-2truecm}
	\plotone{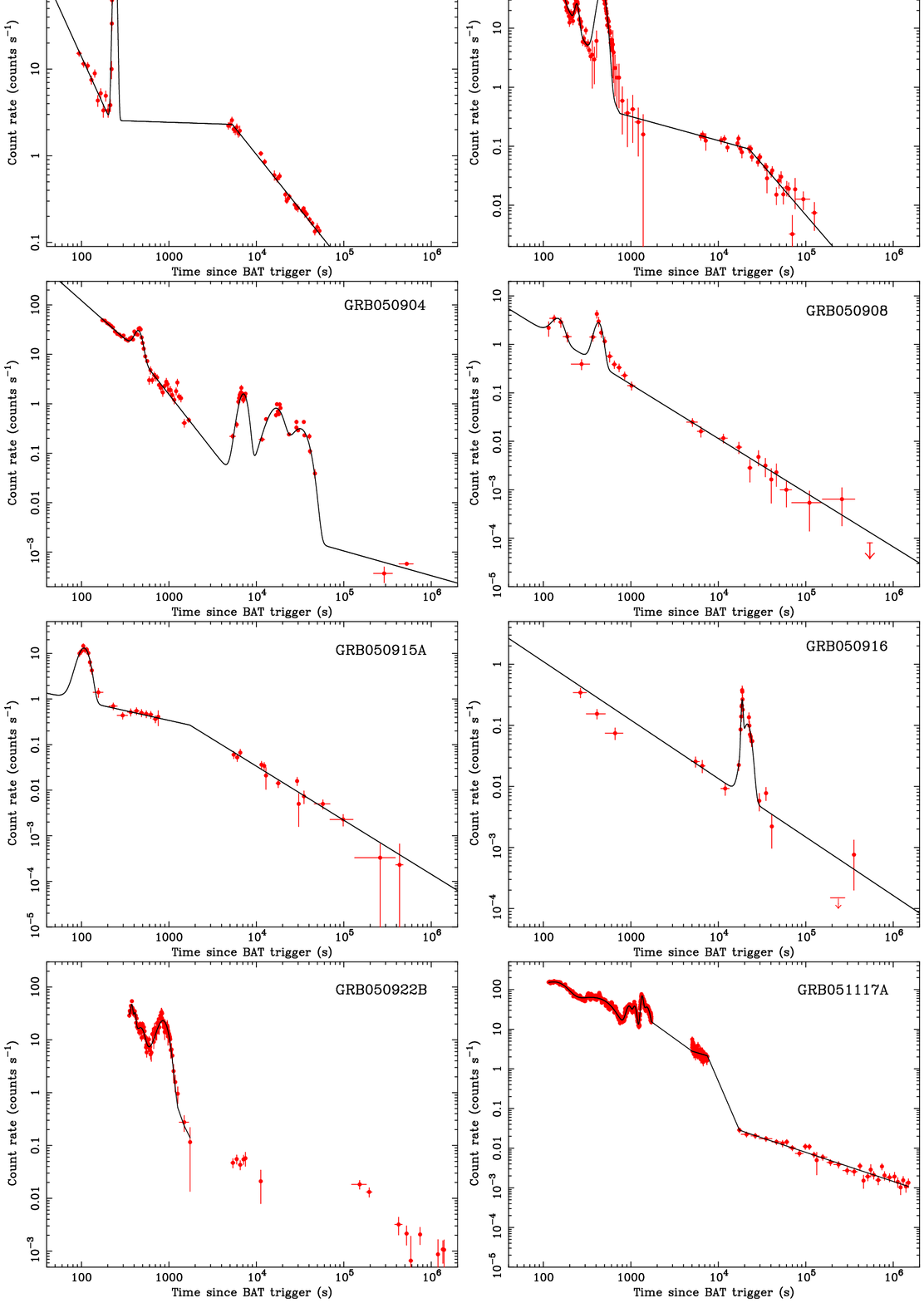}
	\caption[]{Continued.
	\label{flares:bigfig_1c}}
\end{figure*}
\clearpage 
\setcounter{figure}{0}
\begin{figure*}	
	\vspace{-5.5truecm}
	\plotone{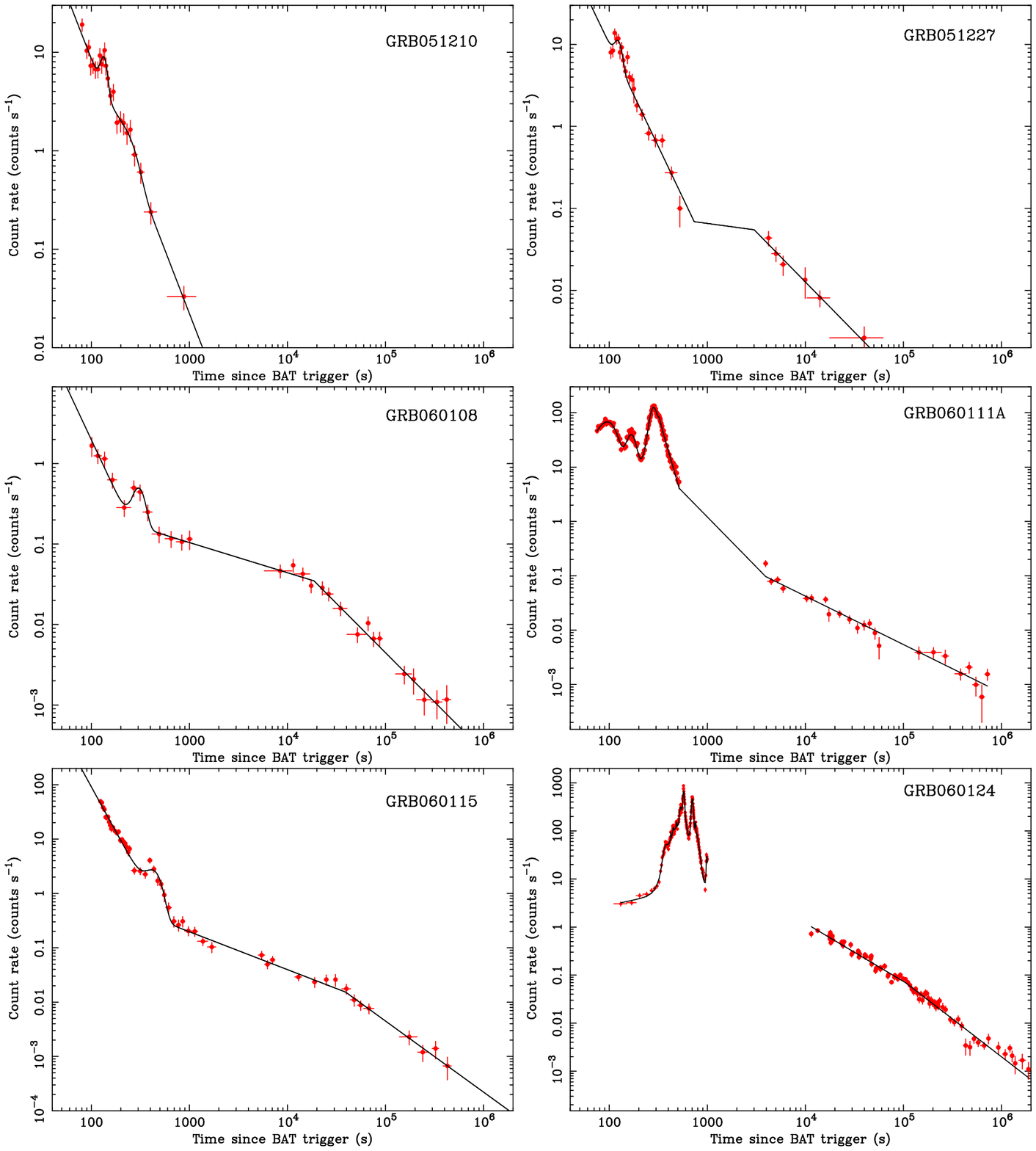}
	\caption[]{Continued.
	\label{flares:bigfig_1d2}}
\end{figure*}

\begin{figure*}[t] 
	\epsscale{2.0}
	\plotone{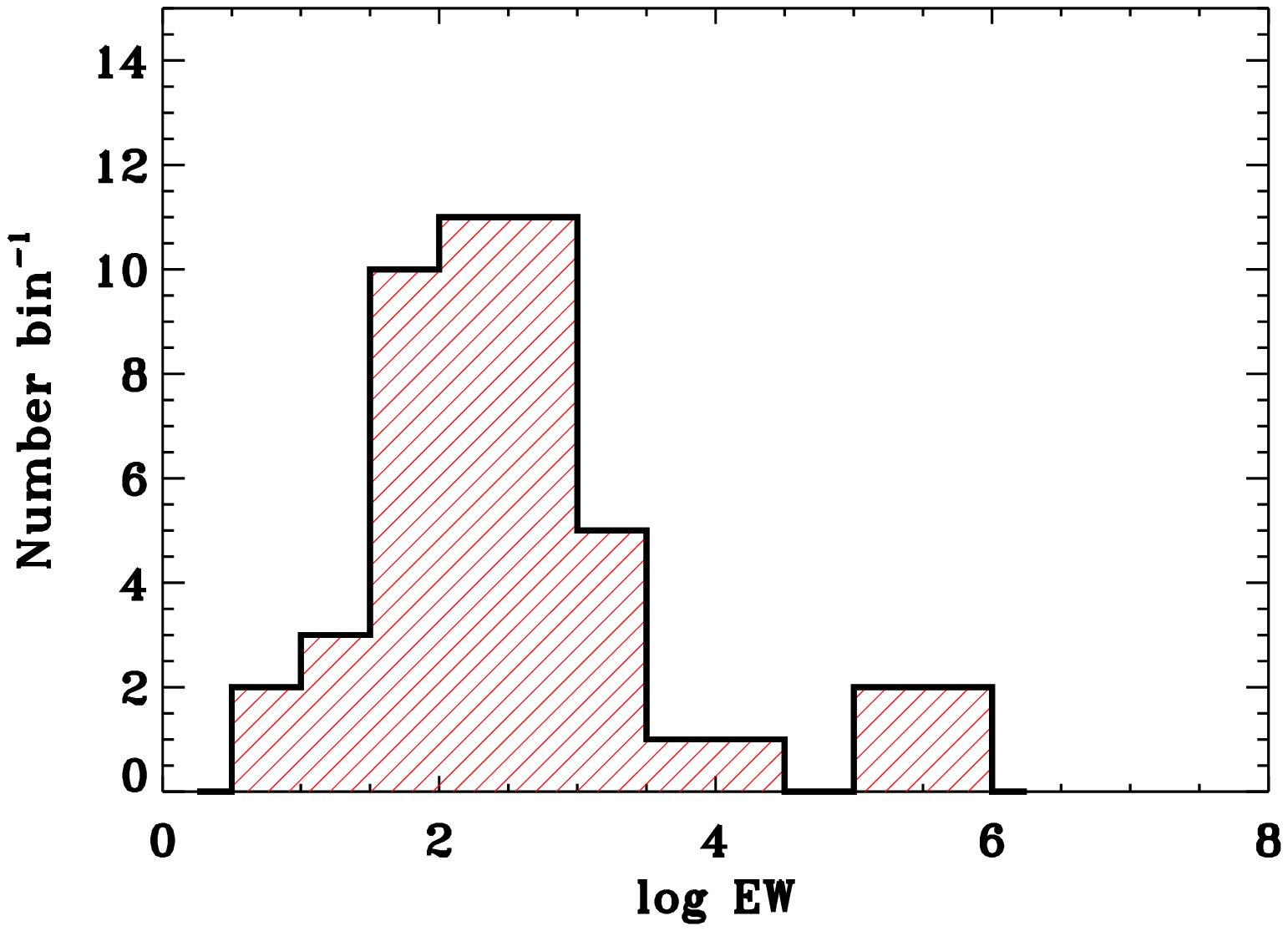}
	\caption[Equivalent width distribution]{Distribution of 
	the equivalent width (in seconds) of the flares. 
	The times are are not corrected for redshift.
	\label{flares:fig:ew_distrib}}
\end{figure*}

\setcounter{figure}{2} 
\begin{figure*}[t]	
	\plotone{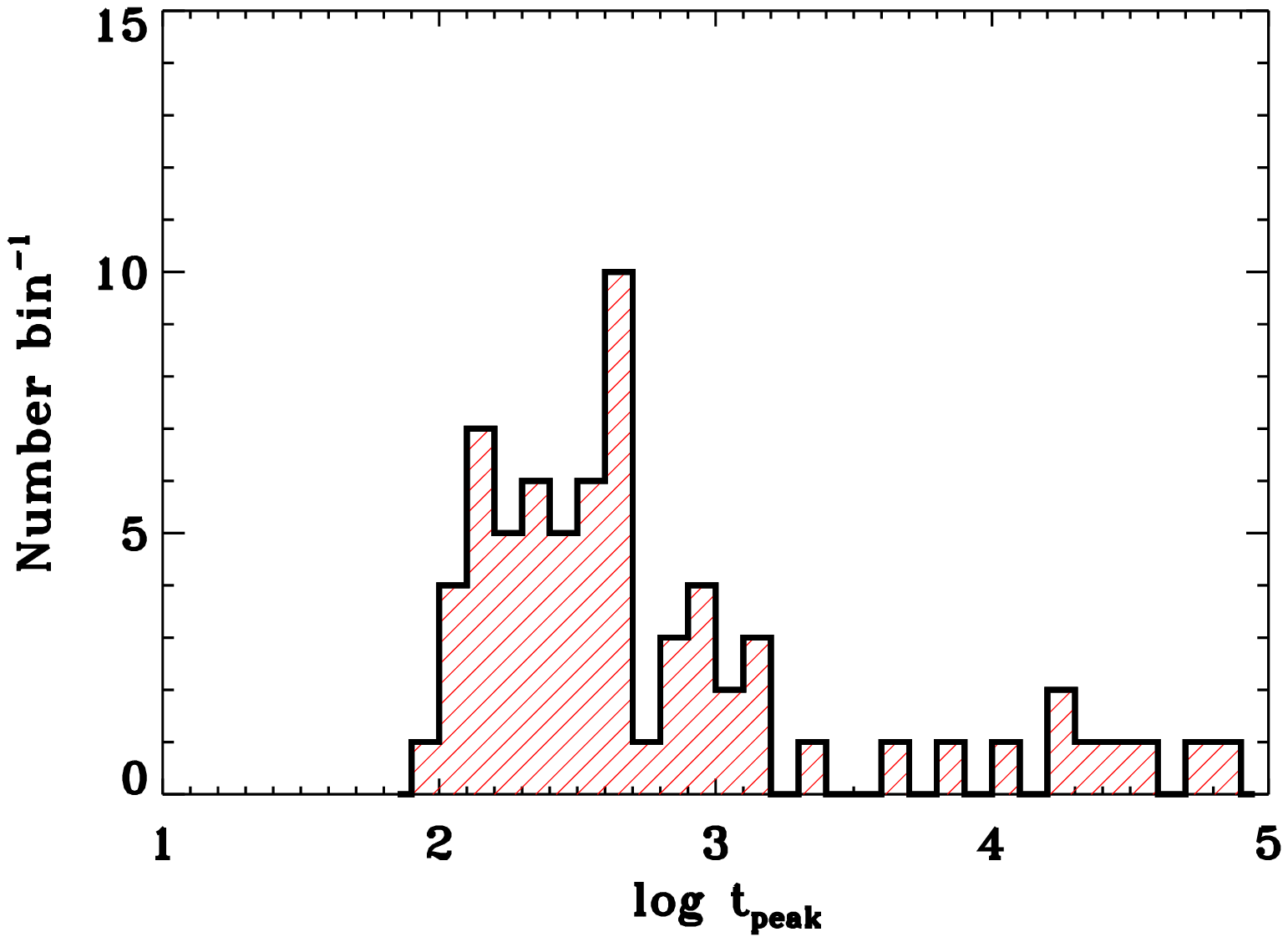}
	\caption[Peak times distribution]{Distribution of the peak times  
	of the flares in excess of the canonical XRT light curve.
	The times are referred to the trigger time, and are not corrected for redshift. 
	\label{flares:fig:peakt_distrib}}
\end{figure*}
\clearpage 

\begin{figure*}[t] 
	\plotone{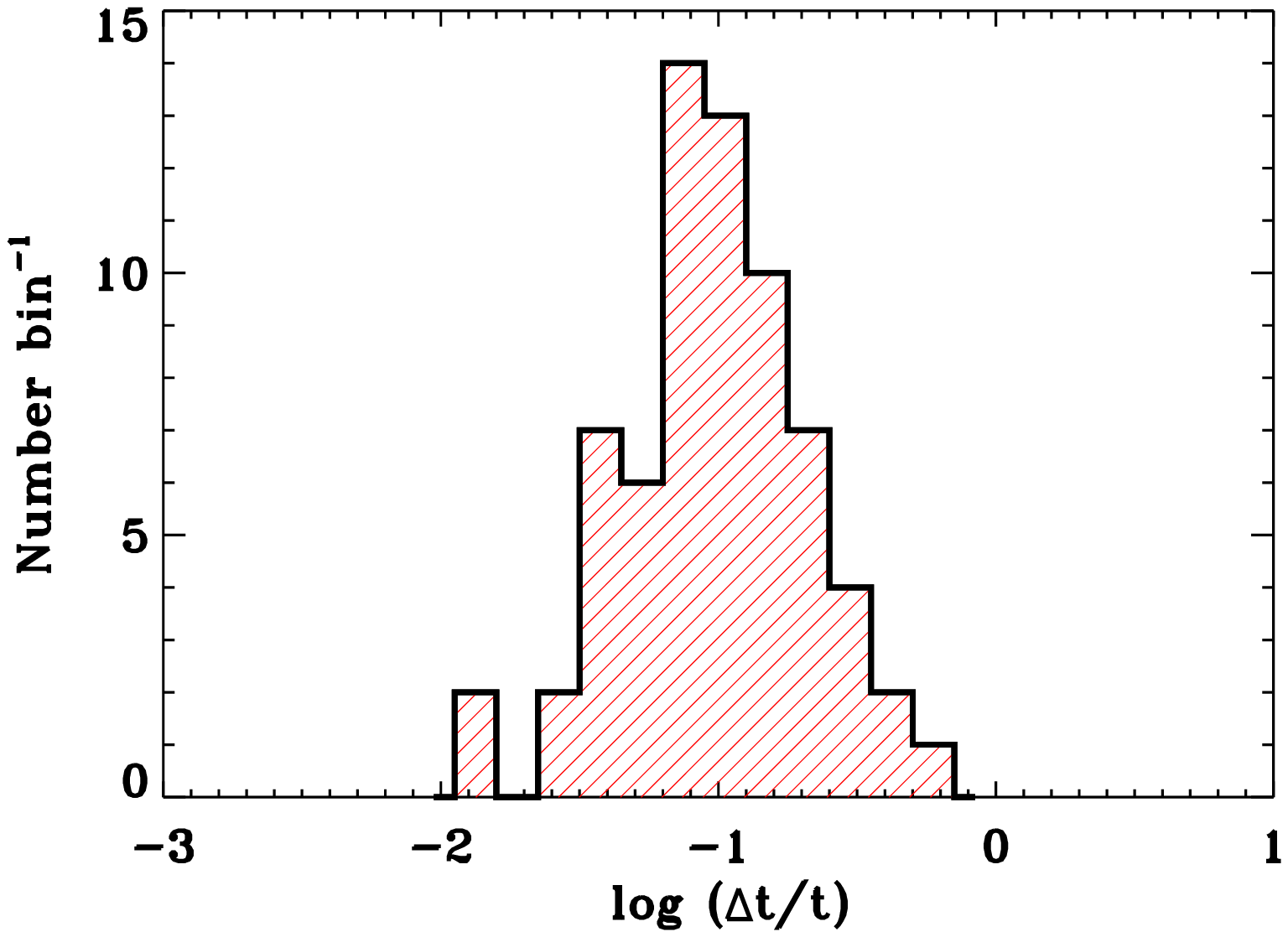}
	\caption[$\Delta t/t$ distribution]{Distribution of the ratio of the flare duration vs.\ 
	the time of occurrence $\Delta t/t$, obtained fitting the flares with 
	Gaussian models (\S\ref{flares:gaussian_fits}), where $\Delta t$ is the width of the Gaussian and 
	$t$ is the Gaussian peak time. 
	This ratio is independent of redshift. 
	\label{flares:fig:dt_t_gauss_distrib}}
\end{figure*}
\clearpage

\begin{figure*}	
	\plotone{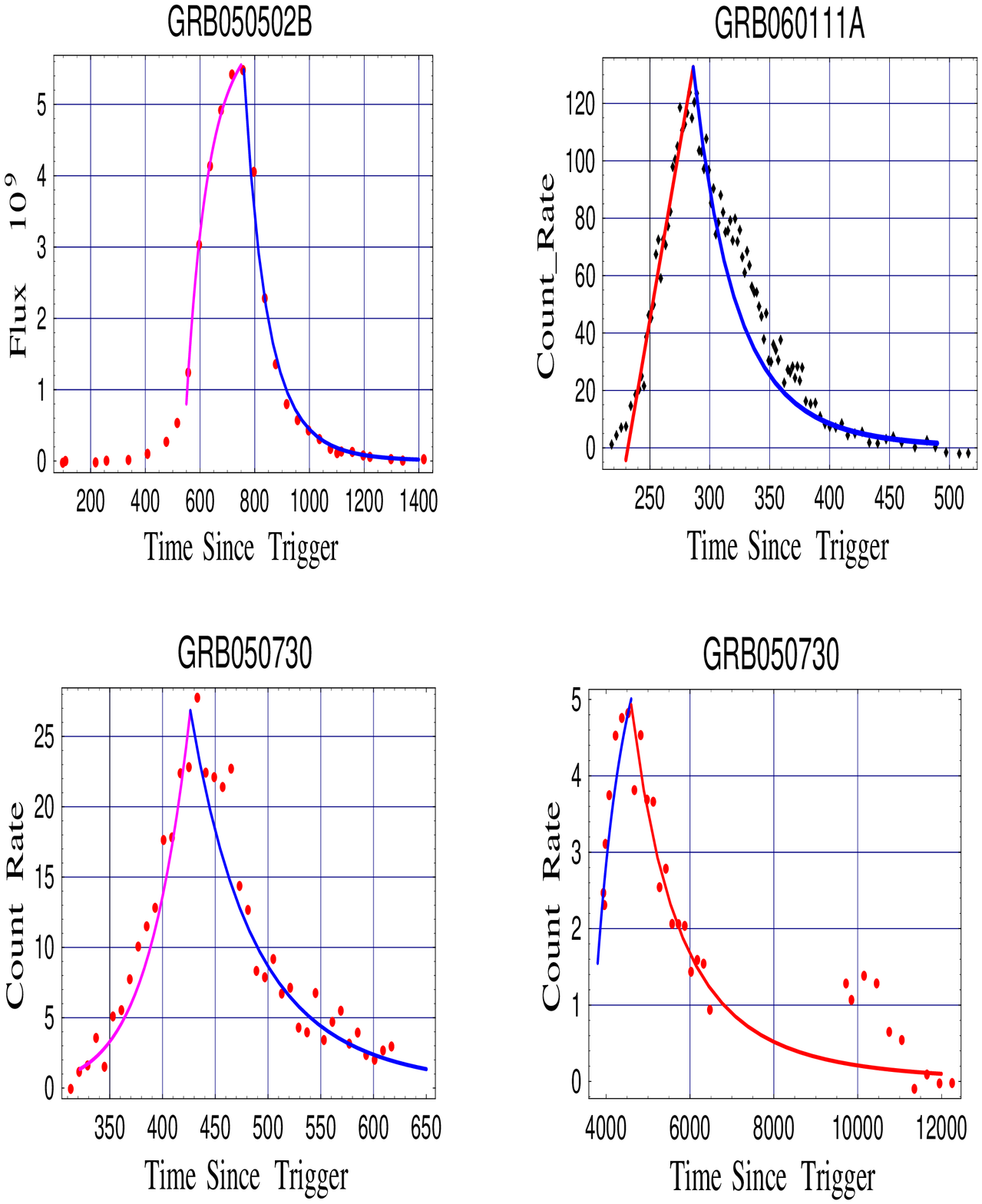}
	\vspace{-2truecm}
	\caption[]{Different flare morphologies as represented by 
	GRB~050502B, GRB~050730, and GRB~060111A. For GRB~050730 different flares are best fit by different
	laws (two power laws for the first, and an exponential rise followed by a power-law decay for the second one.)
	\label{flares:morphology4}}
\end{figure*}
\clearpage 

\begin{figure*}[t] 
	\plotone{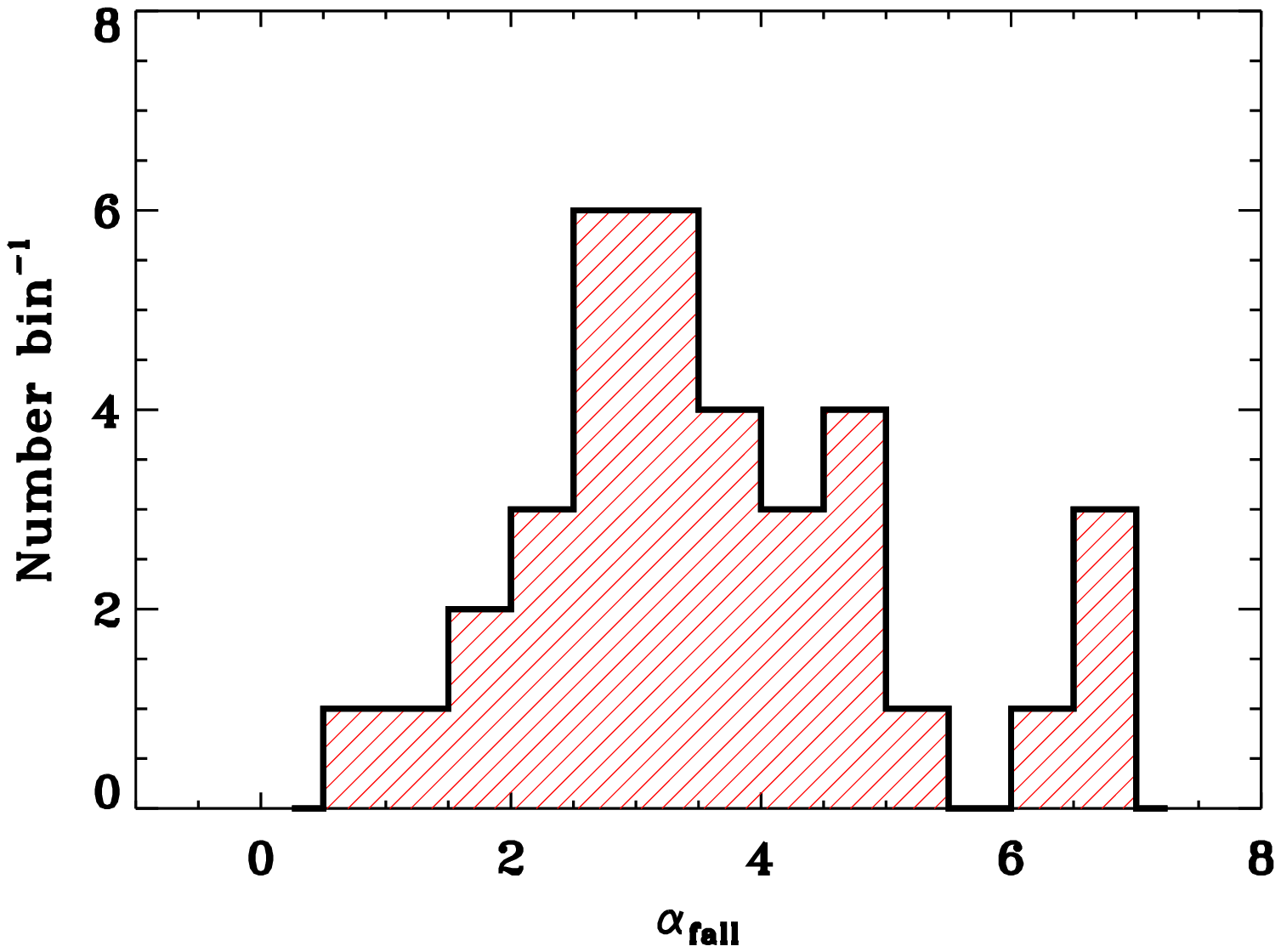}
	\caption[$\alpha_{\rm fall}$ distribution]{Distribution of the decay slope $\alpha_{\rm fall}$ 
	computed using as initial time the point where the flux is 1\% of the peak (see \S\ref{flares:alltaus}). 
	\label{flares:fig:slope_distrib}}
\end{figure*}
\clearpage 

\begin{figure*}[t]	 
	\plotone{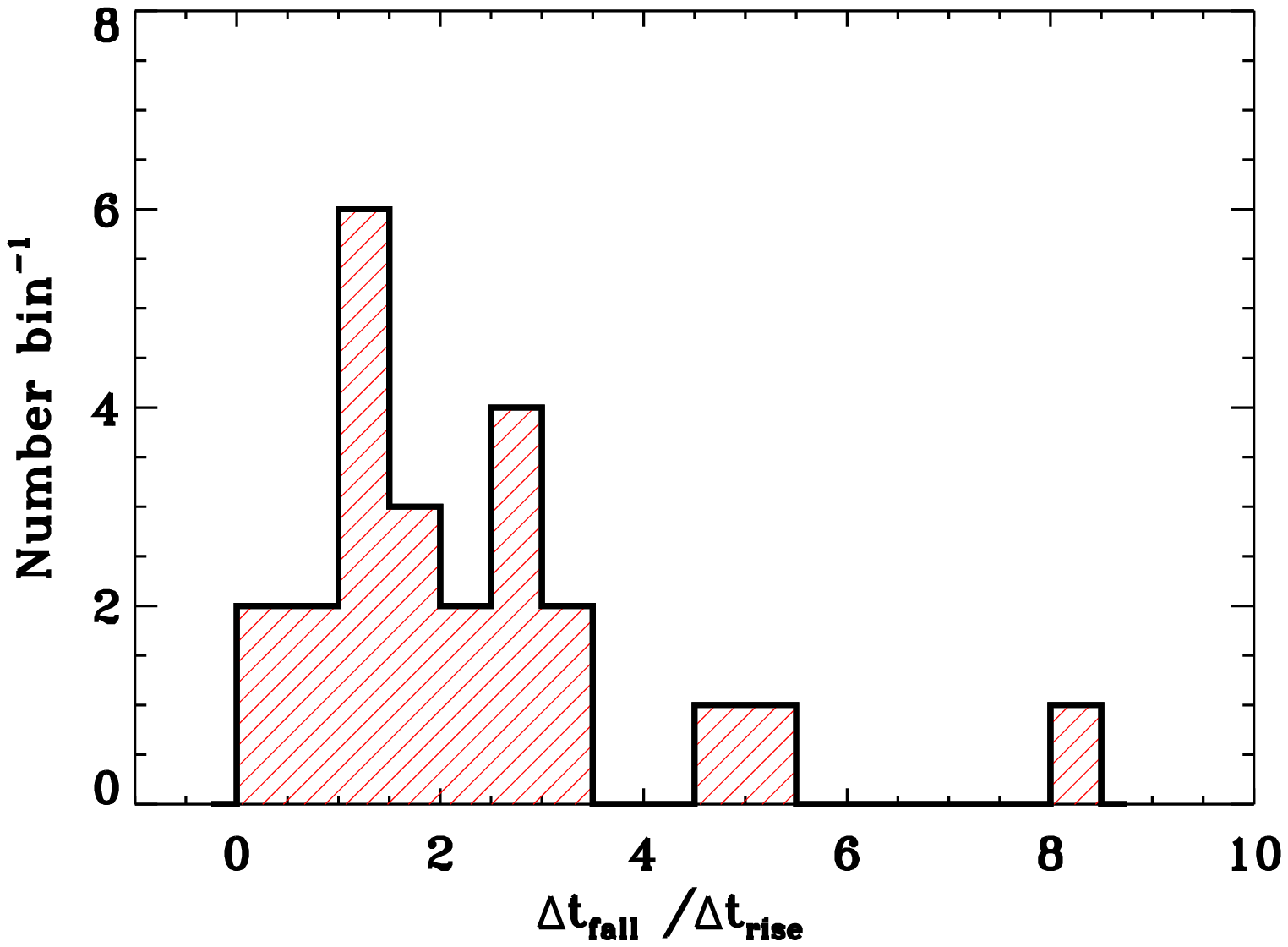}
	\caption[$\Delta t_{\rm fall}/\Delta t_{\rm rise}$ distribution]{Distribution of 
	$\Delta t_{\rm fall}/\Delta t_{\rm rise}$  obtained fitting the flares 
	with power-law and exponential models. 
	These ratios are independent of redshift. 
	\label{flares:fig:rise_fall_distrib}}  
\end{figure*}
\clearpage 

\begin{figure*}[t]      
	\epsscale{2.2}
   \plottwo{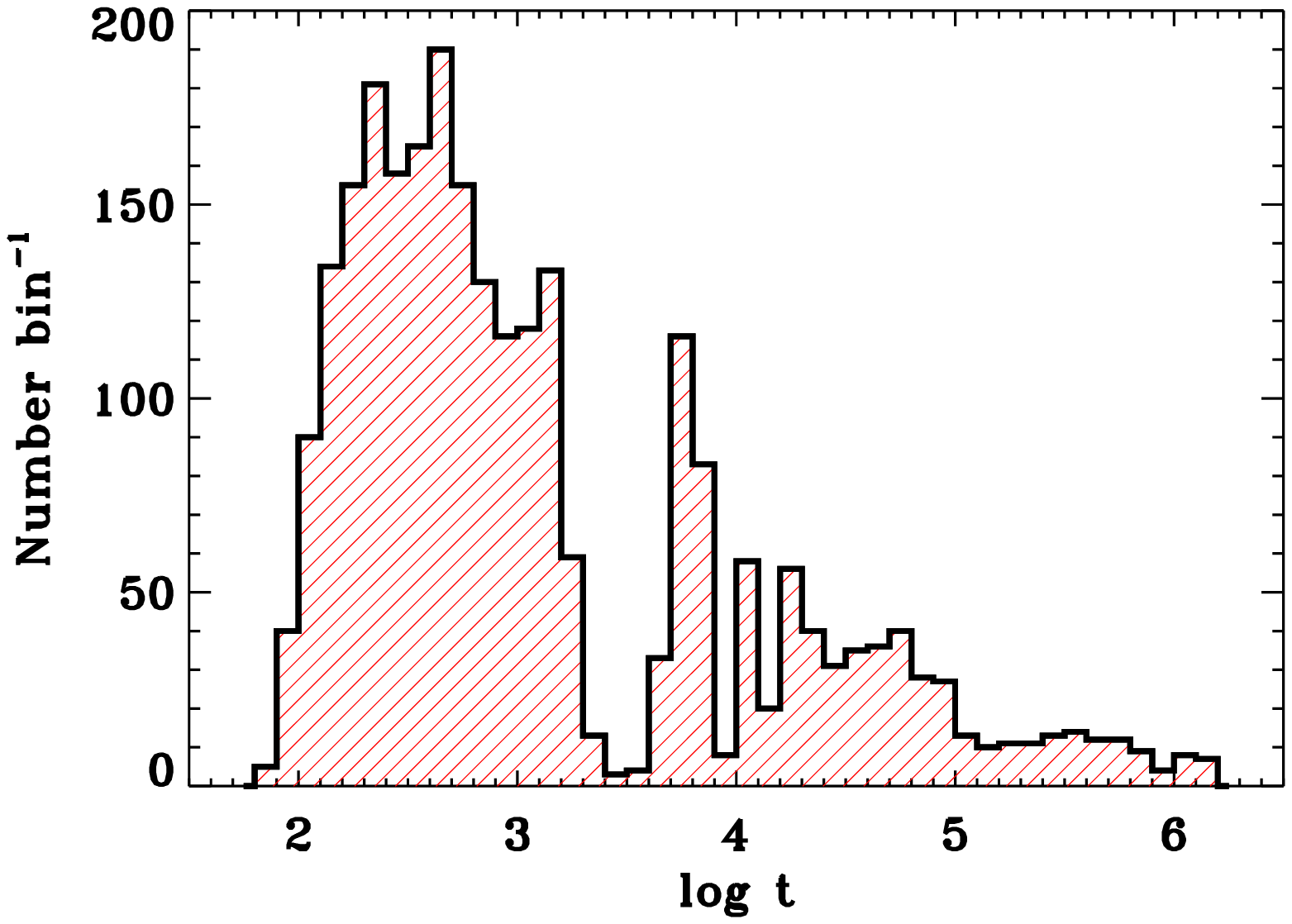}{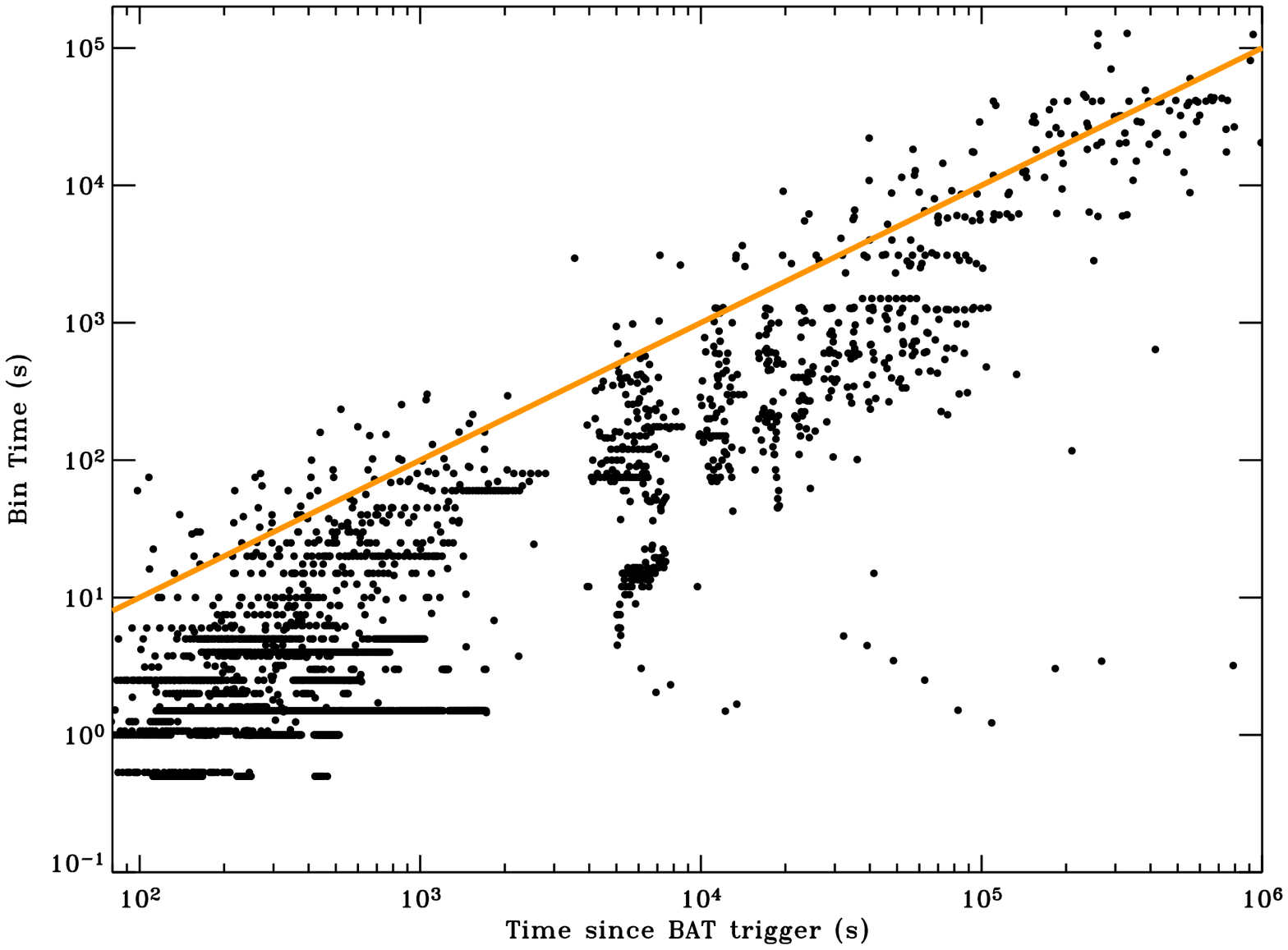}
   \caption[Time resolution vs. time since the BAT trigger.]{
	{\bf (left)}: Distribution of observing times. 
	The gap at $\log t \sim 3.5$ is due to observing constraints (end of the first orbit). 
	{\bf (right)}: Time resolution (BT) as a function of the time since the BAT trigger. 
	The solid curve corresponds to BT$/t=0.1$ and lies above the large majority of the data 
	points.

	\label{flares:fig:bintimes}}
\end{figure*}
\clearpage 

\begin{figure*}[t]      
   \epsscale{2.0}
   \plotone{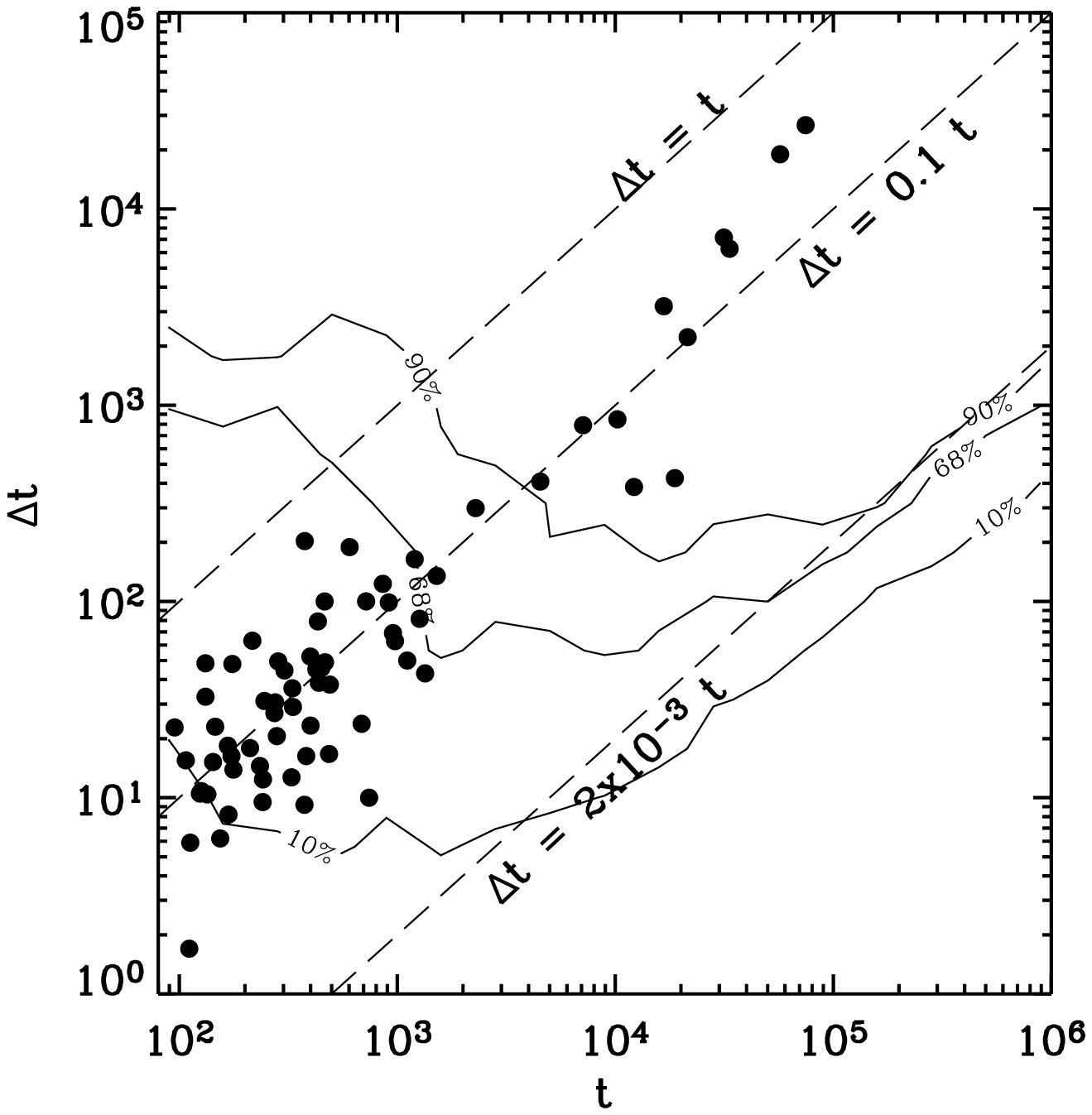}
   \caption[]{The results of the simulation in the $(t, \Delta t)$ plane:
 	    here we plot the contours  for which we have the same detection probability. 
	    Black points are the real data, based on Gaussian widths and peaks. 
		The dashed lines correspond to the three levels 
		$\Delta t = 1$, 0.1, and $2\times 10^{-3}$.  
	    \label{flares:fig:pianosimule}}
\end{figure*}
\clearpage

\begin{figure*}[t]	 
	\plotone{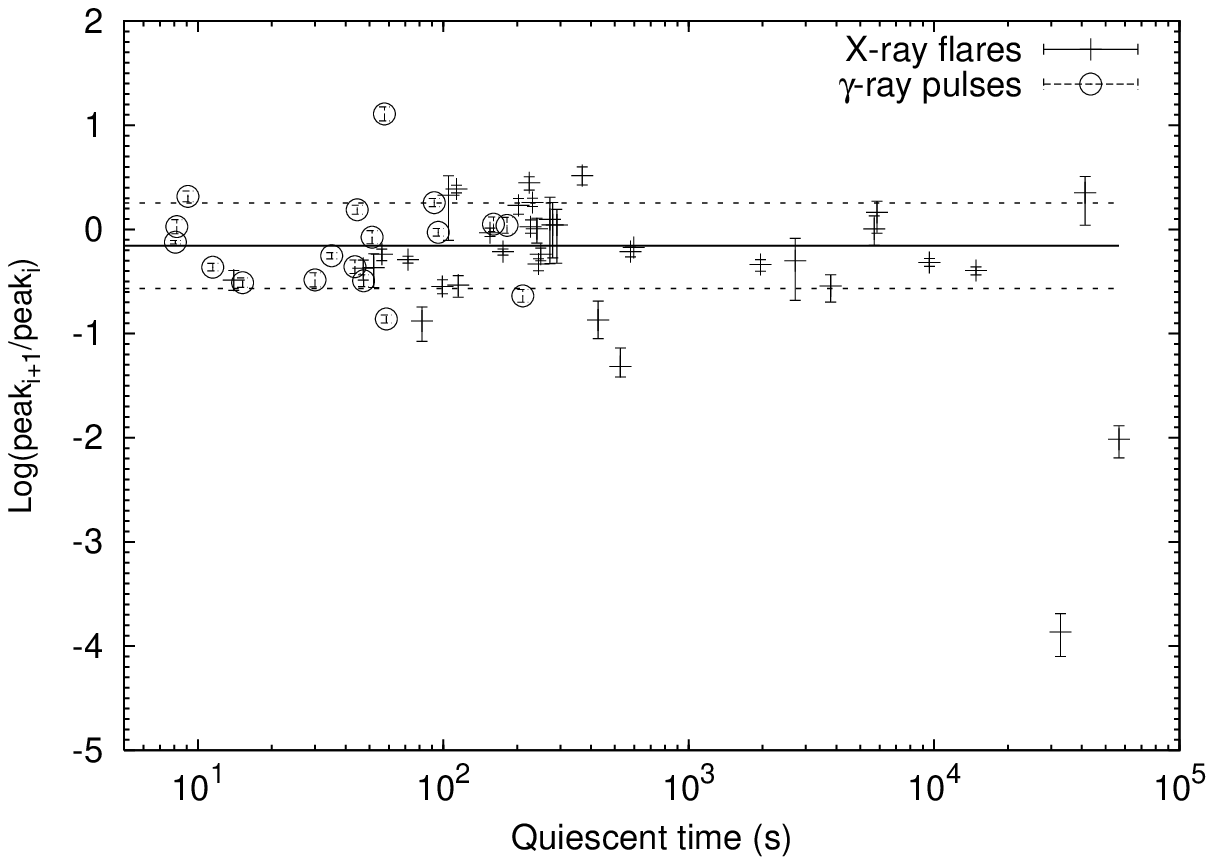}
	\caption{Ratio between the peaks of two successive events for both classes:
	X-ray flares (crosses) and gamma-ray pulses (circles), as a function of
	the quiescent time between the two events. Solid line shows the mean value,
	$-0.157$, when the two points with the lowest ratio are ignored; dashed lines
	show the $\pm1\sigma$ region.
	The outliers are GRB~050724 ($\log {\rm peak}_{i+i}/{\rm peak}_i\sim10^{-2}$)
	and GRB~050502B ($\log {\rm peak}_{i+i}/{\rm peak}_i\sim10^{-4}$).
\label{fig:quiesc_vs_ratio}}
\end{figure*}
\clearpage 

\begin{figure*}[t]	 
	\plotone{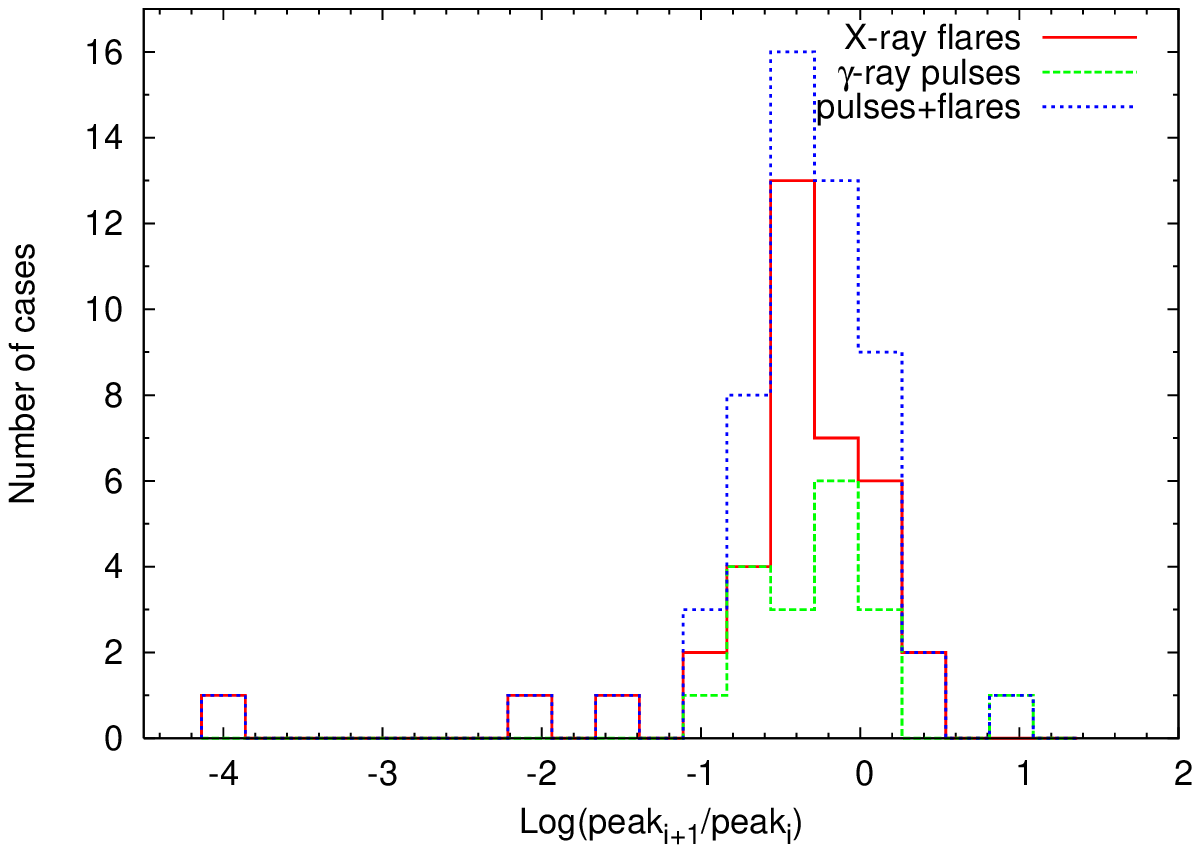}
	\caption{Distribution of the ratio between the peaks of two successive
	events:
	X-ray flares (red), gamma-ray pulses (green), both classes (blue).
	\label{fig:logallratio} }
\end{figure*}
\clearpage 

\begin{figure*}[t]	 
	\plotone{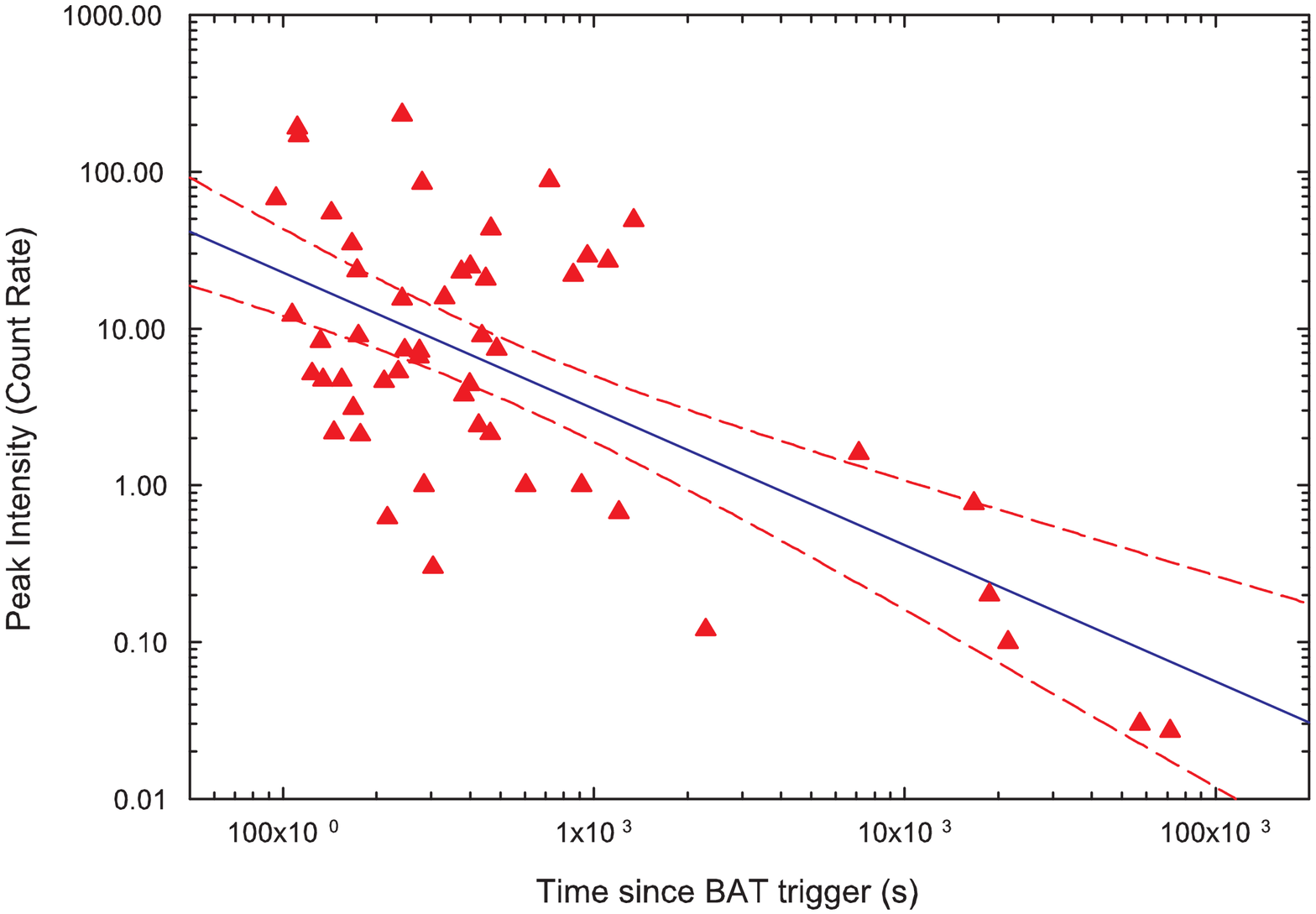}
	\caption[Peak times]{Gaussian peaks of the flares as a function of time. 
	The solid line is the best fit, while the 
	dashed lines correspond to 95\% confidence limits. 
	\label{flares:fig:peakt_norm_corr}}
\end{figure*}
\clearpage 

\begin{figure*}[t] 
	\plotone{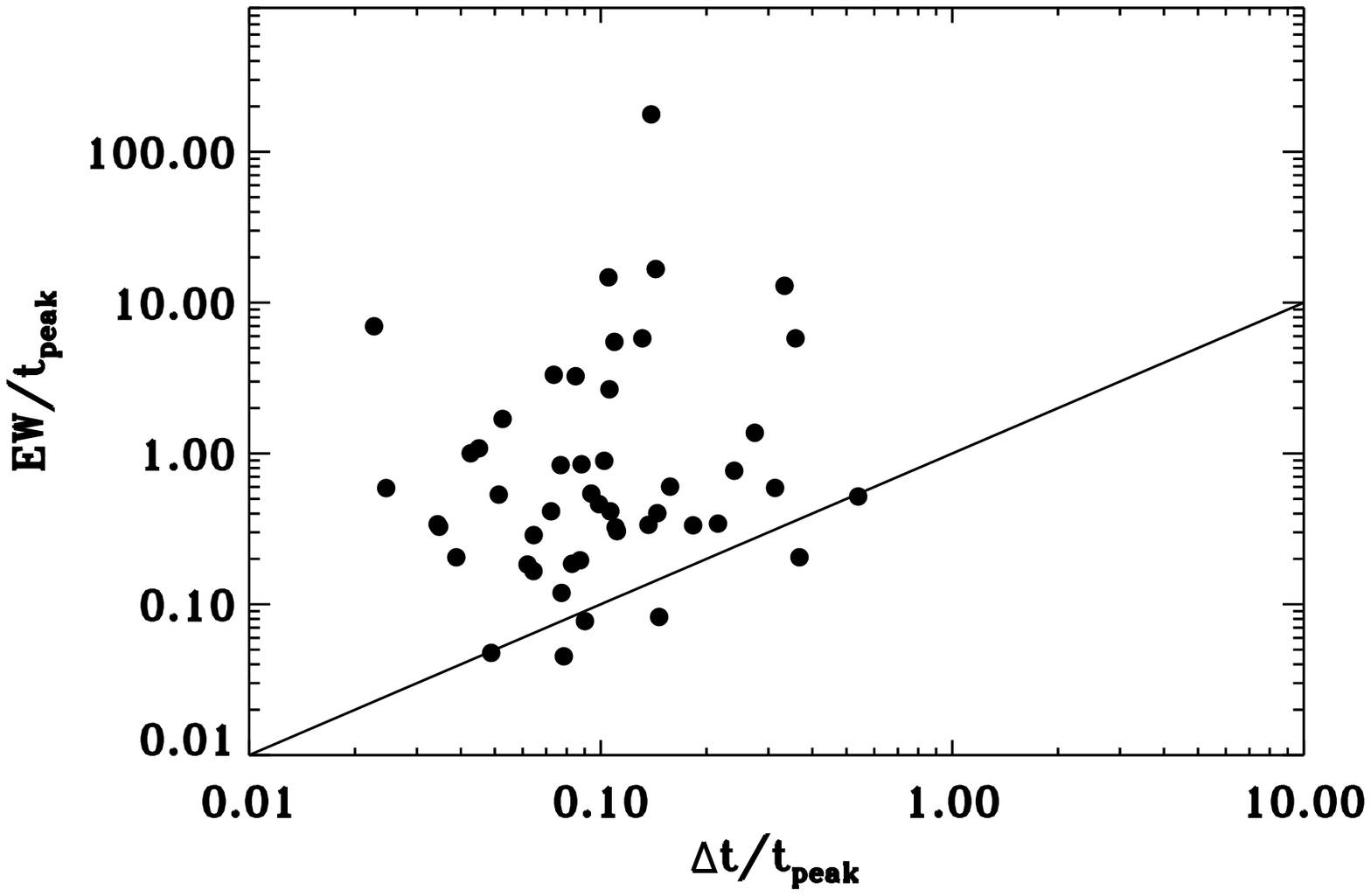}
	\caption[]{$EW/t_{\rm peak}$ vs.\ $\Delta t /t_{\rm peak}$. 
	The solid line is the bisector of the plane. 
	\label{flares:EW_Peak_Time}}
\end{figure*}
\clearpage 

\begin{figure*}[t]	 
	\plotone{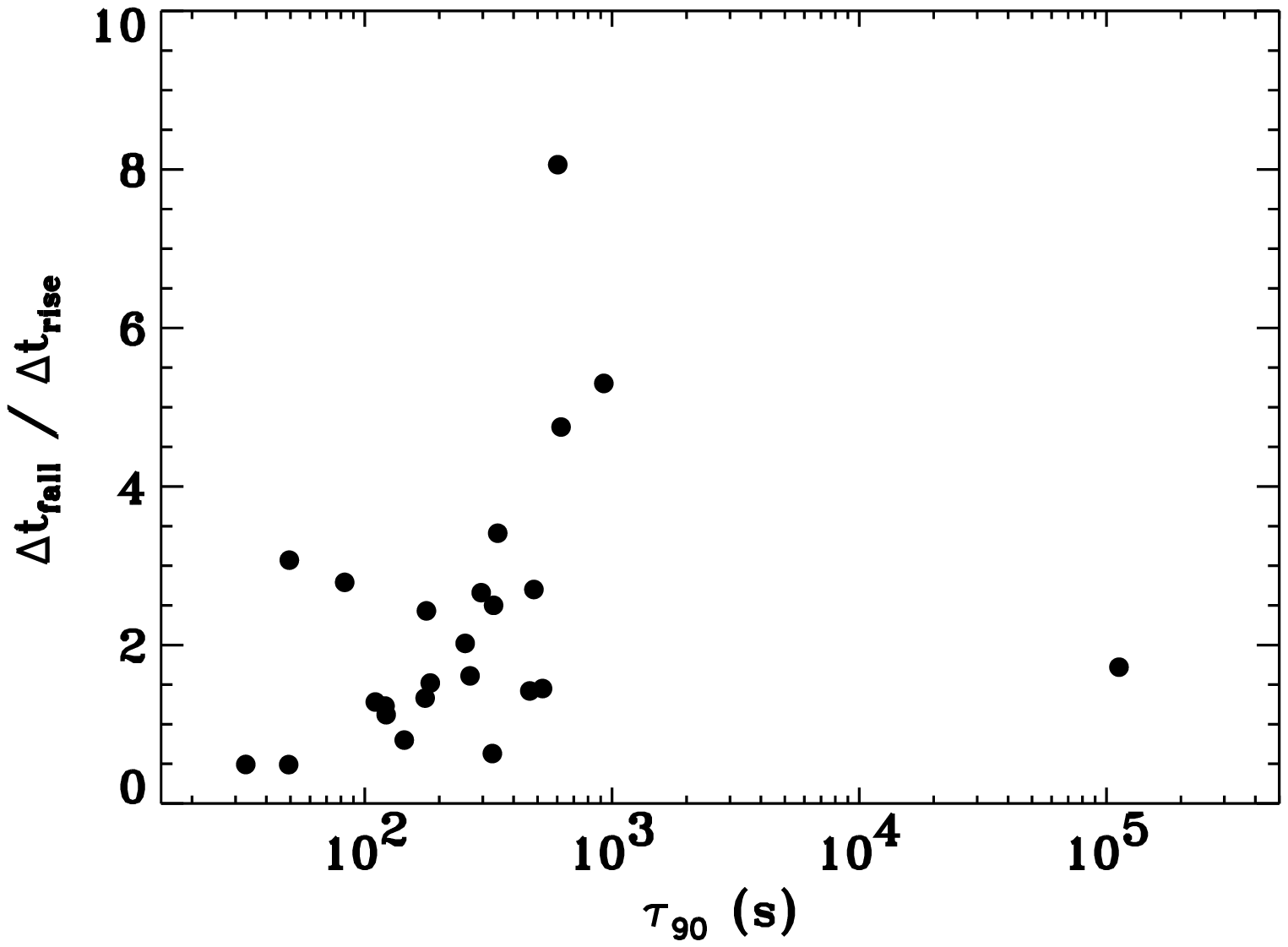}
	\caption[]{$\Delta t_{\rm fall}/\Delta t_{\rm rise}$ vs $\tau_{90}$. 
	\label{flares:taus_t90}}
\end{figure*}
\clearpage

\begin{figure*}[t]	 
	\epsscale{1.8}
	\plotone{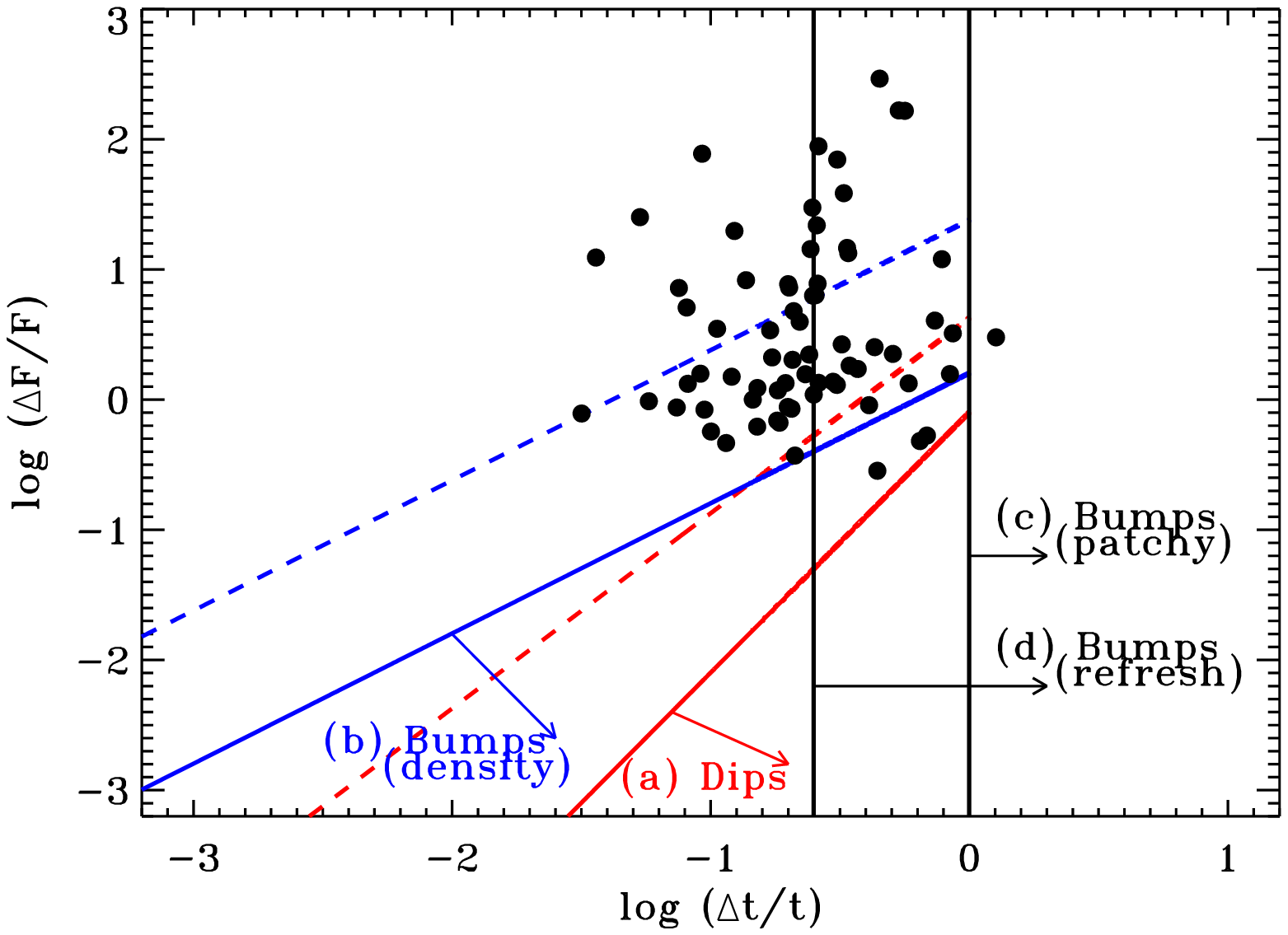}
	\caption[]{Scatter plot of $\Delta F / F$--$\Delta t / t$ values calculated 
	on our flare samples on the kinematically allowed regions for afterglow variabilities 
	according to \citet{Ioka2005:flare_diagnostic}. 
	Data are drawn from Tables  ~\ref{flares:tab:cont_pars} and \ref{flares:tab:fits}: 
	we used the FWHM of the Gaussians as $\Delta t$, the Gaussian peak time for $t$, while
	the ratio of the peak flux over the underlying continuum flux ($\Delta F / F$) was calculated using the
	best fit models.
	The four limits plotted are based on 
	(a) eq.\ (4) in \citet{Ioka2005:flare_diagnostic} for dips (shown on axis), 
	(b) eq.\ (7) for bumps due to density fluctuations (on axis),
	(c) $\Delta t >t$ for bumps due to patchy shells, and 
	(d) $\Delta t > t/4$ for bumps due to refreshed shocks.
	According to \citet{Ioka2005:flare_diagnostic}, when many regions fluctuate 	
	simultaneously, limits a and b are replaced by eqs.\ (A1) and (A2) in 
	\citet{Ioka2005:flare_diagnostic}, respectively.
	The off-axis cases (viewing angle $\theta_v \sim \gamma^{-1}/2 \ga \Delta \theta$, 
	where $\Delta \theta \ga \gamma^{-1}$ is the half-angular size of the variable region) 	
	are shown by dashed lines. 

	\label{flares:yoka}}
\end{figure*}

\end{document}